\documentclass[preprint,review,12pt]{elsarticle}

\usepackage{graphicx}

\usepackage{natbib,amssymb,lineno}

\biboptions{sort&compress}

\journal{International Journal of Heat and Mass Transfer}

\begin{document}

\begin{frontmatter}



\title{Interface thermal behavior in nanomaterials by thermal grating relaxation}


\author{Pier Luca Palla\corref{cor1} }
\cortext[cor1]{Corresponding author}
\ead{pier-luca.palla@isen.iemn.univ-lille1.fr}
\author{Sonia Zampa }
\author{Evelyne Lampin }
\author{Fabrizio Cleri }
\address[label1]{Institut d'Electronique, Microelectronique et Nanotechnologie\\ (IEMN, UMR CNRS 8520),Universit\'e de Lille I, 59652 Villeneuve d'Ascq, France}

\begin{abstract}
We study the relaxation of a thermal grating in multilayer materials with interface thermal resistances. The
analytical development allows for the numerical determination of this thermal
property in Approach to Equilibrium Molecular Dynamics and suggests an
experimental setup for its measurement. Possible non-diffusive effects at the nanoscale are
take into consideration by a non-local formulation of the heat equation. As a case study, we numerically apply the present approach to silicon
grain boundary thermal resistance.

\end{abstract}

\end{frontmatter}



\section{Introduction}\label{secintro}

\begin{figure}
\includegraphics[width=85mm]{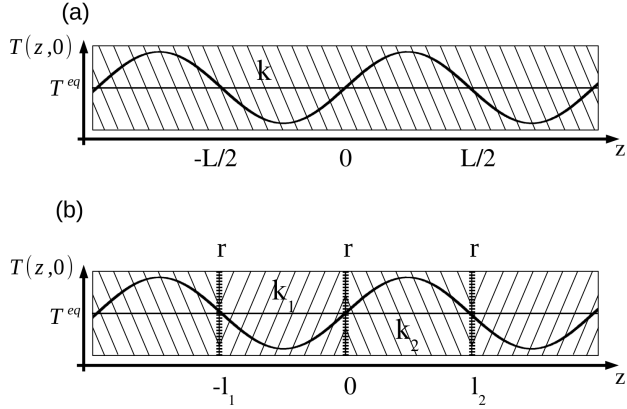}
  \caption{\label{fig:schema1} Schematic representation of a periodic temperature profile (thermal grating) 
in a bulk material of thermal conductivity $k$ (top panel) 
and in bilayer system, of periodicity $L=l_1+l_2$, composed by two materials of conductivity $k_1$ and $k_2$ 
joined by interfaces of thermal resistance $r$ (bottom panel).}
\end{figure}

In this work, we address the heat transport in heterogeneous nanostructured systems.
At small scale, where the surface to volume ratio increases, the behavior
of the interfaces between homogeneous regions is no longer negligible.
Moreover, at least for materials with a large phonon mean free path, also the thermal 
conductivity of the bulk can deviate from the macroscopic picture, i.e. scale effects
can arise. On the other hand, in numerical as in experimental studies, it is often difficult
to distinguish between the bulk and the interfaces role in the thermal response of the overall system.
In order to investigate these issues, we consider here a particular thermal processing, namely the relaxation of 
 a spatially periodic temperature profile, in heterogeneous structures. 
This thermal processing is exploited as well as in numerical techniques, as the Approach to Equilibrium Molecular Dynamics 
(AEMD) \cite{refJAPLampinPalla}, in experimental approaches, as Laser-Induced
Thermal Grating (LITG) \cite{LaserGratingBook}, and in theoretical works
\cite{MaznevPRB84-2011}.
The underlying idea is to generate a periodic temperature
profile or thermal grating (TG) in the system (see Fig. \ref{fig:schema1})
and to observe its equilibration time. This in turn can be related to
the thermal or thermoelastic properties of the media.
Generally speaking, both coherent sound waves and incoherent thermal excitations
are induced in the sample \cite{Landau}. Nevertheless, the most of the energy is
stored in the temperature grating \cite{tempVSwaveTG1,tempVSwaveTG2}. Therefore,
the relaxation is usually described by thermal transport models.
In particular, by considering a classical diffusive thermal transport,
the heat equation predicts an exponential decay of the amplitude of the
TG with a characteristic time $\tau$ related to the thermal conductivity $k$
according to 
\begin{equation}
k = \frac{c L^2}{4 \pi^2  \tau}  \label{eq:kditau0}
\end{equation}
where $L$ is the spatial period of the TG and $c$ the volumetric heat capacity
of the bulk material.

To obtain the temperature profile in LITG experiments, two short laser pulses of central
wavelength $\lambda$ are crossed on a sample resulting in an interference
pattern with period $L =\lambda/2 \sin(\theta/2)$ defined by the angle $\theta$
between the beams. Absorption of laser light leads to a sinusoidal temperature
profile, and the temporal decay of this TG is monitored via diffraction of a
probe laser beam so to measure the bulk conductivity of the medium.
LITG spectroscopy has been widely used in thermal characterizations of materials
\cite{LaserGratingBook} including, e.g.,
silicon membranes \cite{Vega-LI-TTG-advances,LaserGratingBook}, 
and in gas phase diagnostics \cite{LITGexp}.

A similar protocol is implemented in AEMD atomistic simulations. This
numerical technique has been developed in the context of molecular dynamics
with the aim of studying the thermal transport in dielectric materials
\cite{refJAPLampinPalla}. It provides a very efficient and robust method to
investigate thermal properties in a wide range of system sizes.
AEMD has been applied by several authors to the calculation of the
thermal conductivity in c-Si,c-Ge and  $\alpha$-quartz \cite{refJAPLampinPalla},
in Si/Ge nanocomposite \cite{PRL-ClaudioSiGealloy}, in Si nanowires 
\cite{Rurali} and in graphene-based structures \cite{YangChen, TejB}. Initially
developed for classical Molecular Dynamics (MD) this technique has been
recently implemented in ab-initio calculations as well \cite{abinitioAEMDboero}.

These applications of the thermal grating relaxation address the study of
the thermal conductivity in bulk materials. In this paper, with the aim of
extending the AEMD method and of proposing a new experimental setup for thermal
studies, we consider a more general configuration. We address heat conduction in
bilayer systems composed of two homogeneous materials (see Fig.
\ref{fig:schema1}-b), joined through non-ideal interfaces
\cite{giordanopallainterface}. In particular, a resistive interfacial thermal
effect is modeled using a Kapitza resistance \cite{kapitzapaper}.
From the mathematical point of view, this problem represents a further
generalization of the \textit{ Sommerfeld heat-conduction problem for a
ring  } \cite{Sommerfeld}, successively extended by I.R. Vengerov
to the case of two different media simply connected via an ideal interface
\cite{GenSommerfieldRing}.
As we will show in the following, the present development allows for the
theoretical prediction of the ITR in the context of the AEMD. Moreover, the
results suggest that this approach can be adopted in LITG as well.
Therefore, it could provide a direct access to the experimental
determination of the interface thermal resistance (ITR).

This thermal property, usually negligible in macroscopic structures, is
attracting by contrast increasing attention in nanotechnology. For
instance, the removal of the heat generated by electronic systems, one of the
crucial constraint on the performance of modern nanoelectronics, is largely
limited by interfaces between layers. In silicon based thermoelectric
nanomaterials, the grains boundaries resistance allows for a significant
increasing of the figure of merit.

One of the main theoretical limits in projecting down to the nanoscale the 
thermal grating based techniques is the validity of the Fourier's law at such a
scale. We address this crucial point by considering a non-local formulation of
the heat conduction problem. In other words, we introduce in the analytical model
a wavelength dependence of the bulk thermal conductivity. This improvement
perfectly matches the results of the corresponding numerical simulations showing
that, at least down to a given scale, a non-local approach to the thermal
transport allows for a correct description of the ballistic effects in
nanostrucured systems. 

The paper is organized as follows. For the sake of clarity, in Section
\ref{secHeatequation}, we firstly resume the solution of the heat equation for a
periodic bulk system (heat-conduction problem for a ring) and its
application in AEMD simulations \cite{refJAPLampinPalla}. Then, we generalize
the mathematical problem to the bilayer case.
In Section \ref{AEMDresistance}, we discuss the results of numerical
simulations of a bilayer material, namely of a periodic system composed by slabs of crystal silicon (c-Si)
with different crystallographic directions connected by grain boundaries (GB).
The details of the simulations and of the GB model adopted in this work are
reported in \ref{app:simdetails}.
In Section \ref{sec:NonLocalK}, we show that the numerical results
 for the c-Si bulk conductivity \cite{refJAPLampinPalla} can be modeled by a 
 non-local formulation of the heat equation.
 This interpretation is then applied in Section \ref{sec:resfromtau} to the
calculation of the ITR of silicon GB.
Finally, in Section \ref{sec:phase} we discuss the application of the
present developments in LITG experiments, supplying a further analysis 
obtained by additional numerical simulations.

\section{Heat-conduction problem on a ring}\label{secHeatequation}

\subsection{Bulk system}\label{secHeatequationBulk}
We start by considering a mono-dimensional heat conduction problem for an
homogeneous media under periodic boundary conditions (PBC). According to the
classical theory, the heat transfer is governed by the heat equation:
\begin{equation}
c  \frac{\partial T(z,t)}{\partial t} =  k \frac{\partial^2
T(z,t)}{\partial z^2} \label{eq:heateq}
\end{equation}
involving the classical field of temperature $T(z,t)$, and by the
PBC:
\begin{eqnarray}
 T(0,t)&=&T(L,t)\label{pbcbulk1} \\
 \frac{\partial T(0,t) }{ \partial z} &=& \frac{\partial T(L,t) }{
\partial z} \label{pbcbulk2}
\end{eqnarray}
Eq. (\ref{pbcbulk1}) states the periodicity of the solution and, in particular, 
the continuity of the temperature field across the periodic boundaries.
Eq. (\ref{pbcbulk2}) represents the continuity of the heat flux according to the Fourier's Law
\begin{equation}
 J(z,t)= -k\frac{\partial T(z,t)}{\partial z} \label{eq:Fourier}
\end{equation}
or, equivalently, the conservation of the energy.

For an arbitrary initial condition $T(z,0)$ $z \in
[-\frac{L}{2},\frac{L}{2}]$,
the well-known solution of the heat equation is  
\begin{equation}
T(z,t)= T^{eq} +\sum_{n=1}^{\infty} \left[ A_n cos(\alpha^0_n z) +  B_n
sin(\alpha^0_n z) \right] e^{-\frac{t}{\tau^0_n}} \label{solutionbulk}
\end{equation}
where $T^{eq}$ is the asymptotically target temperature and
\begin{equation}
 \tau^0_n = \frac{c}{k (\alpha^0_n)^2} \label{eq:taualpha1}
\end{equation}
The PBC in Eqs. (\ref{pbcbulk1}) and (\ref{pbcbulk2}) impose the following set
of wavenumbers:
\begin{eqnarray}
\alpha^0_n &=& n \frac{2\pi}{L} \label{eq:alphaFourier}
\end{eqnarray}
The coefficients $A_n$ and $B_n$ are the components of the Fourier series
of $T(z,0)$. Eq. (\ref{solutionbulk}) shows that each term in the series is
damped by an exponential factor with characteristic time $\tau^0_n$.
Combining Eq. (\ref{eq:alphaFourier}) with Eq. (\ref{eq:taualpha1}), we easily
get
\begin{eqnarray}
\tau^0_n &=&  \frac{c L^2}{4 \pi^2 k} \frac{1}{n^2} \label{tau0}
\end{eqnarray}

In LITG experiments, a sinusoidal TG of wavenumber $\alpha^0_1$ is initially
induced in the sample.
Therefore the solution reads
\begin{equation}
T(z,t) = T^{eq} + A \sin(\alpha^0_1 z)  e^{-\frac{t}{\tau^0_1}} \label{solutionbulksin}
\end{equation}

On the other hand, in AEMD simulations, the initial TG is usually a step-like
profile with one half of the simulation box at temperature $T_1$ and the other
half at $T_2$. 
Therefore, the relaxation is not rigorously mono-exponential. Nevertheless,
being $\tau^0_n \propto \frac{1}{n^2}$, the higher the order of the
harmonic, the shorter its life time and we can approximate the relaxation
through a single exponential factor, as discussed below.

In order to describe the thermal equilibration of the periodic profile,
in AEMD calculations we need to consider only the difference $\Delta T(t)$
between the average temperatures of the two halves of the system.
This difference of temperature reads
\begin{eqnarray}
\Delta T(t) &=& \frac{2}{L}\int^{\frac{L}{2}}_{0} T(z,t)dz - 
\frac{2}{L}\int^0_{-\frac{L}{2}} T(z,t)dz \nonumber \\
&=& \sum_{n=1}^{\infty} \phi_n  e^{-\frac{t}{\tau^0_n}}     \label{eq:difftemp}
\end{eqnarray}
The coefficients $\phi_n$ are proportional to $A_n/n$ (or $B_n/n$) so that,
for a step-like initial temperature profile, we have $\phi_n \propto
\frac{1}{n^2}$ as well.
Moreover, by considering the expression in Eq. (\ref{solutionbulk}), we can state that
only the terms with an even value of $n$ give a contribution to the summation. Hence,
the second non-zero term in $\Delta T(t)$ has a characteristic time
$\tau^0_3$ nine times smaller than the fundamental time $\tau^0_1$. 
In conclusion, the TG becomes rapidly sinusoidal and the equilibration is
matter-of-factly mono-exponential, i.e. after a brief transient the solution in
Eq. (\ref{solutionbulksin}) is recovered.

In practice, knowledge of the leading characteristic time $\tau^0_1$
allows for the determination of the thermal conductivity via Eq.
(\ref{tau0}) with $n=1$, i.e. via Eq. (\ref{eq:kditau0}).

\subsection{Bilayer system with ITR}\label{secHeatequationMultiL}

In this section, we generalize the development of
Section \ref{secHeatequationBulk} to the case of the bilayer structure shown
in Fig. \ref{fig:schema1}-b. A mono-exponential behavior, similar to the bulk
case, is recovered and the decay time is related to the thermal response of the
interface, i.e. to the ITR.

We define the temperature fields $T^{(1)}(z,t)$ and $T^{(2)}(z,t)$ in the
two homogeneous regions of length $l_1$ and $l_2$ and conductivity
$k_1$ and $k_2$ respectively and we consider a mono dimensional heat equation
for each region:
\begin{equation}
c_i  \frac{\partial T^{(i)}}{\partial t} =  k_i  \frac{\partial^2
T^{(i)}}{\partial z^2} \label{eq:heateqmulti}
\end{equation}
for ${i = 1,2}$, $c_i$ being the volumetric heat capacity of the two materials. 
According to Kapitza resistance definition, a discontinuity in the temperature field
$\delta T =  r J$ is introduced at the interfaces.
Therefore, Eq. (\ref{eq:heateqmulti}) is solved with the following set
of PBC:
\begin{eqnarray}
 T_1(0,t)-T_2(0,t) &=& r J(0,t)\label{eq:pbcGB1nonloc} \\
 T_2(l_2,t)-T_1(-l_1,t) &=& r J_1(-l_1,t) \label{eq:pbcGB2nonloc} \\
 J_1(0,t) &=& J_2(0,t) \label{eq:pbcGB3nonloc}\\
 J_1(-l_1,t) &=&J_2(l_2,t)  \label{eq:pbcGB4nonloc}
\end{eqnarray}
where the flux $J_i$ are related to the temperature field via Eq.
(\ref{eq:Fourier}).
The general solution of Eq. (\ref{eq:heateqmulti}) for an arbitrary initial
condition reads:
\begin{equation}
T^{(i)}(z,t) = T^{eq} + \sum_{n=1}^{\infty} \theta_n^{(i)} (t) \chi_n^{(i)}(z) \label{eq:gensol}
\end{equation}
where
\begin{eqnarray}
\theta^{(i)}_n(t) &=& e^{-\frac{t}{\tau^{(i)}_n}} 
\\
\chi^{(i)}_n(z) &=& a^{(i)}_n \, cos(\alpha^{(i)}_n z ) + b^{(i)}_n \,
sin(\alpha^{(i)}_n z )
\end{eqnarray}
and
\begin{equation}
\tau^{(i)}_n = \frac{c_i}{k_i (\alpha^{(i)}_n)^2}
\label{eq:taudef}
\end{equation}
The allowed wave numbers $\alpha_n^{(i)}$ are eventually obtained by imposing
the PBC (\ref{eq:pbcGB1nonloc}),
(\ref{eq:pbcGB2nonloc}), (\ref{eq:pbcGB3nonloc}), and  (\ref{eq:pbcGB4nonloc}). 
The sinusoidal functions  $\chi^{(i)}_n(z)$ represent a complete orthonormal
basis for the solution of the present problem. As shown in Section
\ref{secHeatequationBulk}, in the case of a homogeneous system these
functions turn out to be the well-known Fourier basis (being $\alpha_n = n
\frac{2\pi}{L}$). 
$a^{(i)}_n$ and $b^{(i)}_n$ are the coefficients of the initial
condition in this basis.
Moreover, PBC requires that 
\begin{equation}
\tau^{(1)}_n = \tau^{(2)}_n=\tau_n \label{eq:taucons}
\end{equation}
This leads to the following relation between the wavenumbers:
\begin{equation}
 (\alpha^{(1)}_n) ^2  \frac{k_1}{c_1} = (\alpha^{(2)}_n)^2  \frac{k_2}{c_2}
\end{equation}

The complete solution is again a superposition of periodic
functions of the position damped by exponential functions of the time. Each term of
the superposition has a different decay time related to the corresponding 
wave number by Eq. (\ref{eq:taudef}).

For the sake of simplicity, we consider the same bulk properties in the right
side and in the left side of the interface (the general case is exposed in
\ref{app:nonlocal}) and we focus on the interface effects due to the ITR.
So, we assume
\begin{eqnarray}
c_1&=&c_2=c  \\
k_1&=&k_2=k
\end{eqnarray}
hence, the sets of wavenumbers in the two homogeneous regions
coincide
\begin{eqnarray}
\alpha^{(1)}_n  = \alpha^{(2)}_n =\alpha_n 
\end{eqnarray}
Moreover, we impose the condition
\begin{eqnarray}
l_1 = l_2 = l =\frac{L}{2}
\end{eqnarray}

The PBC provide a set of linear equations for the coefficients $a^{(i)}_n$ and
$b^{(i)}_n$ involved in the definition of the basis functions $\chi^{(i)}_n(z)$.
By imposing that the determinant of this set of equations is zero, we obtain
the following transcendental equation for the wavenumber $\alpha_n$
\begin{eqnarray}
 \left( \cos(2 l \alpha_n)-1 \right) (r k \alpha_n)^2  + 4\sin(2 l \alpha_n) r k
\alpha_n \nonumber \\ - 4\left( \cos(2 l \alpha_n)-1 \right)  = 0 \label{eq:det}
\end{eqnarray}
with the $\tau_n$ related to the $\alpha_n$ via Eq. (\ref{eq:taudef}), i.e.
\begin{eqnarray}
\alpha_n = \sqrt{\frac{c}{k \tau_n}}\label{eq:alphaTau}
\end{eqnarray}
It is not possible to solve this equation analytically to find the allowed
values of $\alpha_n$ (or $\tau_n$), therefore we cannot write the analytical solution $T^{(i)}(z,t)$.
Nevertheless, via Eq. (\ref{eq:det}), the ITR $r$ can be expressed as a function of
the $\alpha_n$ according to
\begin{eqnarray}
r = 2\frac{ \sin(2l\alpha_n) + \sqrt{ 2(1-\cos(2l\alpha_n))
}}{(1-\cos(2l\alpha_n)) k \alpha_n }
\label{eq:r1r2} 
\end{eqnarray}
This means that knowledge of a single $\alpha_n$ or equivalently of one of
the decay times $\tau_n$ allows for the calculation of $r$.
Eq. (\ref{eq:r1r2}) represents, indeed, the main analytical result of
the present work.

In the following, we discuss the set of wavenumbers involved in the
present problem and we show that, also in the bilayer case, only the longest 
decay time drives the equilibration of the TG. 

First of all, we note that a sequence of solutions of Eq. (\ref{eq:det}) are
obtained for
\begin{eqnarray}
\alpha^0_n =\frac{\pi n}{l}
\end{eqnarray}
or, in terms of the decay times, for
\begin{eqnarray}
\tau^0_n= \frac{c l^2}{k \pi^2} \frac{1}{n^2} \label{eq:tau0inNonhomo}
\end{eqnarray}
 These solutions are independent of the value of $r$ and coincide with the
decay times (\ref{tau0}) of the bulk system.
As a matter of fact, they describe the thermal equilibration of components 
of the initial condition that doesn't require a heat conduction across the interfaces.
In other words, their independence of the interface resistance $r$ implies that they only represent
situations where the heat flux across the interfaces is zero. The
series of $\tau^0_n$ coincides indeed with the decay times of a system of length $L$
under adiabatic boundary conditions (no heat transfer at boundaries).

\begin{figure}
\begin{center}
\includegraphics[width=80mm]{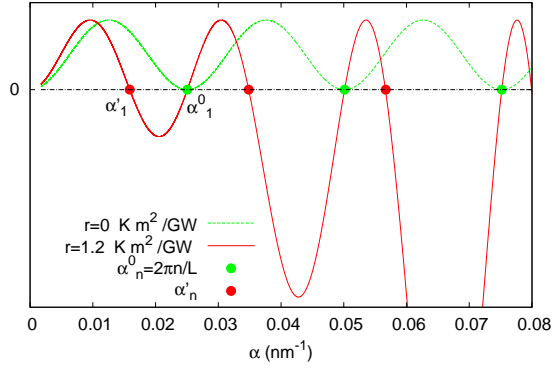}
\end{center}
\caption{\label{fig:detVSAlpha} Wavenumbers, obtained via Eq. (\ref{eq:det}), involved in 
the solution of the heat conduction problem for a bilayer system in Eq. (\ref{eq:gensol}).
The curves are the left-hand side of Eq. (\ref{eq:det}) with $l = L/2 = 125$ nm, $k = 68$ W/mK 
in the cases $r=0$ (bulk) and $r=1.2$ m$^2K$/GW (bilayer).
This system is representative of the results obtained in the following sections
for the case study considered in the present work, namely a c-Si bilayer system
with (100)$\Sigma 29$ GBs.}
\end{figure}

In Fig. \ref{fig:detVSAlpha}, we report the left-hand side of Eq. (\ref{eq:det})
as a function of $\alpha$ for the bulk system, i.e. $r=0$, and for $r=
1.2$ m$^2$K/GW, the intercepts with the horizontal axes supply the
corresponding wavenumbers.
For $r=0$ the sequence $\alpha^0_n = \frac{2 \pi n}{L}$ are correctly recovered.
As $r$ increases a different sequence of wavenumbers, $\alpha'_n$, is
obtained, partially coincident with the $r=0$ set.
Therefore, in the bilayer case ($r \neq 0$) the set of basis function
${\chi^{(i)}_n(z)}$ may be split into two subspaces. The first one is composed
by sinusoidal functions with wavenumber
$\alpha^0_n$ independent of $r$ and characterized by an heat flux equal to zero
at the interfaces.
Being the heat flux proportional to the space derivative of the temperature, this subspace is
composed by {\it even } sinusoidal functions.
The second subspace, characterized by the wavenumbers ${\alpha'_n}$, 
is therefore composed by {\it odd} sinusoidal functions.

Hence, for the odd-parity initial condition here considered (see Fig.
\ref{fig:schema1}) only the ${\alpha'_n}$ wavenumbers appear in the solution in
Eq.(\ref{eq:gensol}). Consequently, only the decay times 
\begin{equation}
\tau'_n =  \frac{c}{k (\alpha'_n)^2}  
\end{equation}
are considered in the following. 
The case of an arbitrary initial condition will be discussed in Section \ref{sec:phase}.

\begin{figure}[ht]
\begin{center}
\includegraphics[width=80mm]{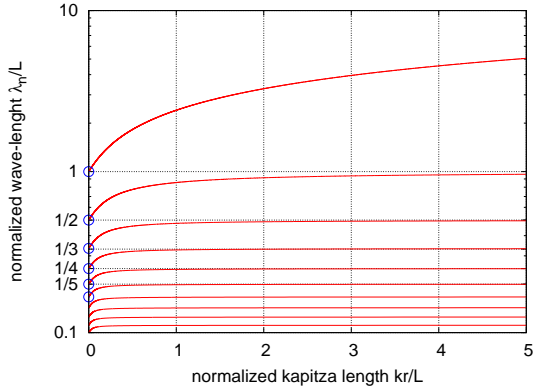}
\end{center}
\caption{\label{fig:kapitzashift} Normalized wavelengths of the base functions for 
the solution of the heat equation (\ref{eq:heateqmulti}) as a function of the normalized Kapitza 
length (inverse of the Biot number) in the case $l_1=l_2$ and $k_1=k_2$.
}  
\end{figure}

The set of wavenumbers $\alpha'_n$ can be numerically studied, and a good deal
of physical insight is gained by defining the normalized wavelength
$\frac{\lambda_n}{L}$ and the {\it normalized Kapitza length} \cite{Barrat}
$\frac{k r }{L}$.  With these variables, the equation (\ref{eq:det}) turns into the
universal function (i.e. independent of the specific parameters of the
problem) plotted in Fig. \ref{fig:kapitzashift}.
If the Kapitza length is zero, i.e. in the bulk case where $r=0$, the Fourier
basis wavelengths are recovered, namely $\lambda_n=\frac{L}{n}$.
On the other hand, the introduction of the periodic pattern of ITR induces a
progressive spread of these wavelengths. In particular, in the limit $rk/L
\rightarrow \infty$, $\lambda_1$ tents to infinity while the others, for $n>1$,
are limited and tent to $\frac{L}{n-1}$. By considering that the decay time
$\tau_n$ of each component is proportional to $\lambda_n^2$, the wavelengths
with $n>1$ turn to be negligible in this limit.
Therefore, for high values of the Kapitza length the temperature profile in the
bulk regions is flat and the {\it lumped hypothesis} can be applied.
Indeed the normalized Kapitza length corresponds to the inverse of the Biot
number, therefore in the $rk/L \rightarrow \infty$ limit the small Biot number
regime is recovered.

\begin{figure}
\begin{center}
\includegraphics[width=80mm]{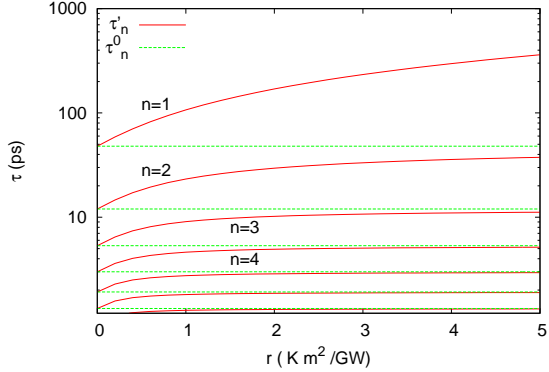}
\end{center}
\caption{\label{fig:tauVSr} Decay times $\tau_n$ of the TG
defined in Eqs. (\ref{eq:gensol}) and (\ref{eq:det}) for a bilayer silicon
system with period $L = 250$ nm and thermal conductivity $k=68$ W/mK. In
particular we plot the solutions of Eq. (\ref{eq:det}) as a
function of the ITR $r$.  }
\end{figure}

In Fig. \ref{fig:tauVSr}, we plot the decay times $\tau'_n$ actually involved in Eq.
(\ref{eq:gensol}) as a function of the ITR $r$.
Clearly, $\tau'_n > \tau^0_n = \tau^0_1 / n^2 $ for all
$n$ and, moreover, for each value of $n$ the difference  $\tau'_n - \tau^0_n$ is a
decreasing function of $n$.
Therefore, we can state that also in the present bilayer case just
one exponential decay times, namely $\tau'_1$, drives the system equilibration, being largely
dominant on to the others.

In conclusion, this leading decay time, eventually measured in experiments or
obtained by AEMD simulations, allows to extract the value of the ITR by
using Eqs. (\ref{eq:r1r2}) and (\ref{eq:alphaTau}) providing that the bulk
conductivity $k$ of the bulk regions is known.

\section{Numerical simulations of a bilayer system}\label{AEMDresistance}

  As a straightforward application of the results reported in the previous
section, we have applied the AEMD method to the calculation of the interface
thermal resistance of a grain boundary (GB) in c-Si.
  To this aim, the periodic structure in Fig. \ref{fig:schema1}(bottom) has been
realized by means of two crystalline slides of the same thickness ($l_1=l_2=l$)
with different crystallographic orientations joined by a grain boundary. The
two slides exhibit the same thermal conductivity ($k_1=k_2=k$). 
  Periodic boundary conditions are applied in all the directions, therefore the
system is actually composed by two semi infinite layers, parallel to the $x-y$
plane, and two crystallographically identical GBs are present in the supercell. 
In \ref{app:simdetails}, we report the details of the simulations and
the analysis of the atomistic structure of the GBs.

\begin{figure}
\begin{center}
\includegraphics[width=80mm]{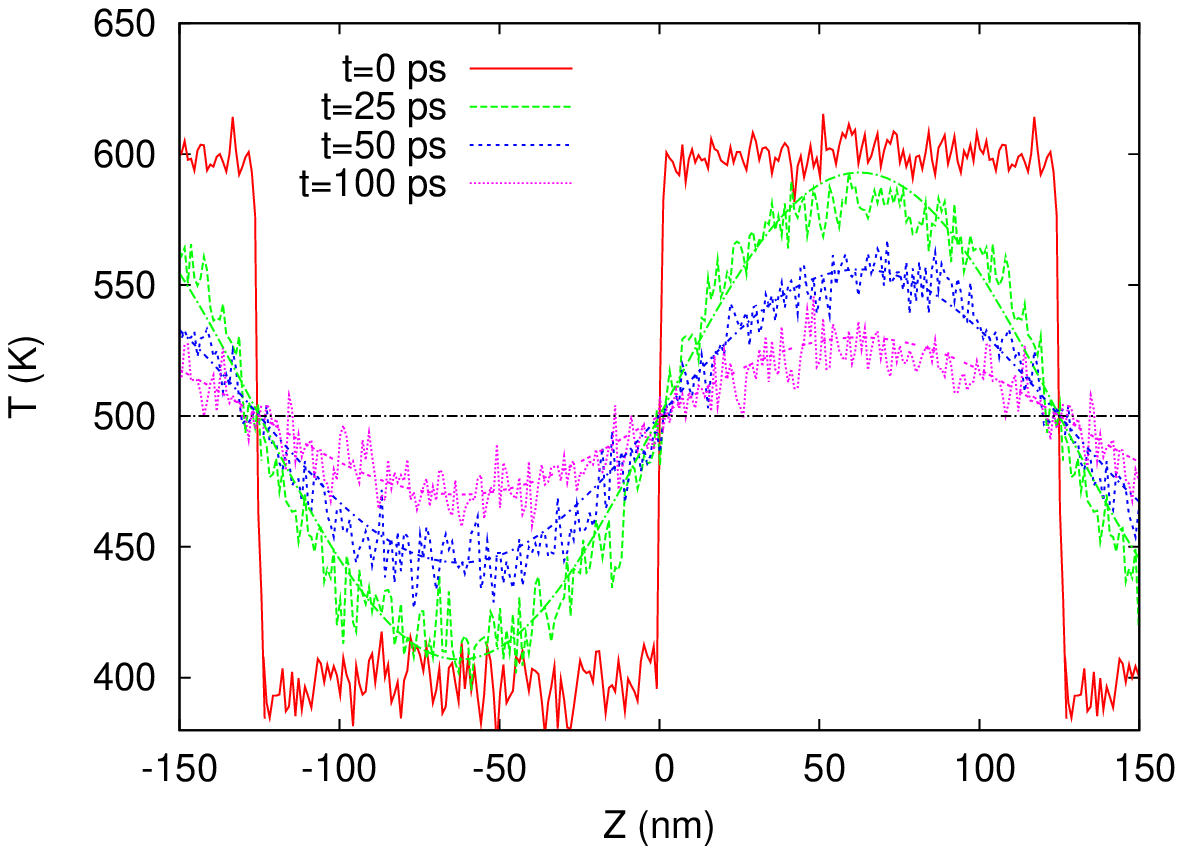}
\includegraphics[width=80mm]{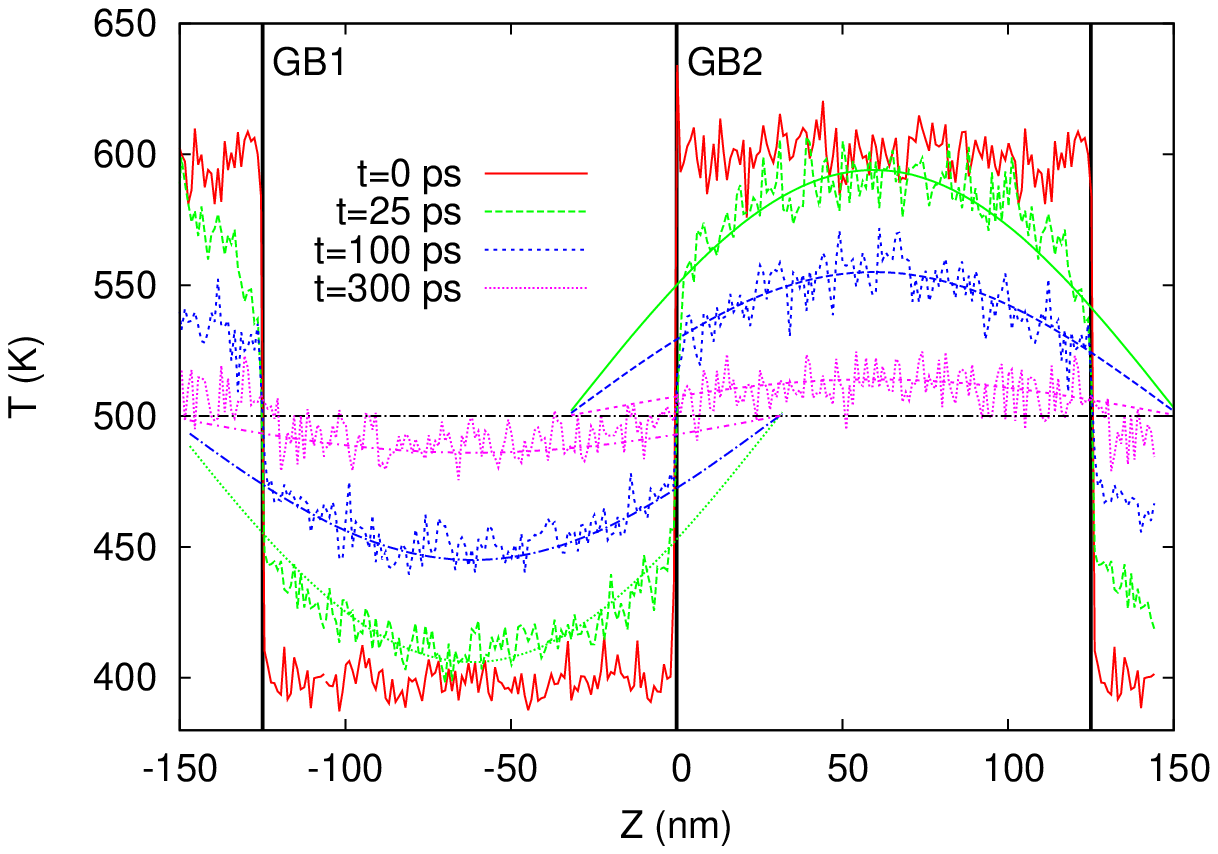}
\end{center}
\caption{Temperature profiles in a bulk (top panel) and in a
bilayer system (bottom panel) during AEMD simulations. 
\label{fig:tempprof} }
\end{figure}

 According to the AEMD method, the two halves of these systems have been heated
at two different temperature $T_1=400$ K and $T_2=600$ K so to obtain a
initial step-like periodic temperature profile along the z-direction. Afterward,
we have let the system equilibrate via a microcanonical simulation.
  The typical evolution of the TG during the equilibration
is plotted in the bottom panel of Fig. \ref{fig:tempprof} while, in the top
panel, we show the bulk case for comparison. In both cases, the initial
step-like profile rapidly turns into a smooth sinusoidal curve in the two
homogeneous regions. In presence of GBs at the interfaces, a jump of the
temperature is observed due to their ITR. For the bulk system,
the wavenumber of the sinusoidal profile coincides with $\alpha_1^0=\frac{2
\pi}{L}$. 
In the bilayer case, the profile exhibits a larger wavelength 
$\lambda_1=\frac{2 \pi}{\alpha_1}$. Coherently with the discussion
presented in the previous section, after a brief transient:
\begin{equation}
T^{(i)}(z,t) \simeq T^{eq} +  b^{(i)}_1
\sin(\alpha'_1 z) e^{-\frac{t}{\tau'_1}} \label{solutionmulti}
\end{equation}

\begin{figure}
\begin{center}
\includegraphics[width=80mm]{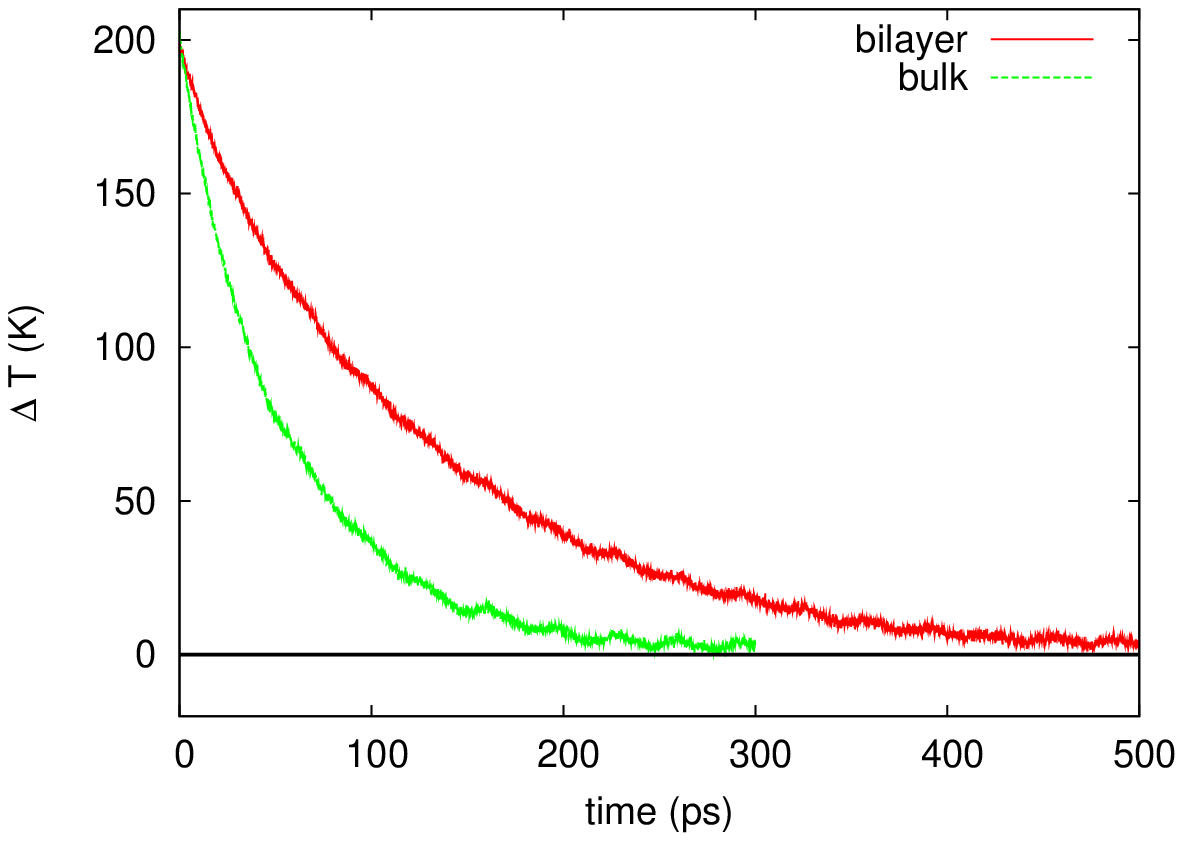}
\caption{\label{fig:tempdec} Temporal evolution of the difference of the average temperatures in the 
two homogeneous regions. In the initial step-like profile, the two regions are equilibrated at $T1=$400 K 
and $T2=$600 K.
}
\end{center}
\end{figure}

\begin{figure}
\begin{center}
\includegraphics[width=80mm]{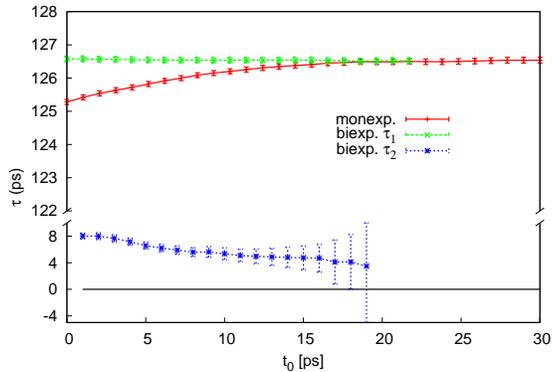}
\caption{\label{fig:Taufit} Analysis of the relaxation of a TG in a silicon bilayer system.
The simulation results in Fig. \ref{fig:tempdec} are fitted on a mono-exponential function and on a bi-exponential function.
The corresponding characteristic times are plotted as function of the left
endpoint, $t_0$, of the fitting interval. 
}
\end{center}
\end{figure}

In Fig.\ref{fig:tempdec}, we plot  the difference between the
average temperatures of the two halves of the systems as a function of the time.
In order to exclude the initial multi-exponential behavior and to extract
the leading decay time from such a curve, the following fitting procedure
have been applied:
  a mono-exponential and a bi-exponential function have been fitted to
the $\Delta T(t)$ data in the range $(t_0, t_{max})$, $t_{max}$ being the
instant when
$\Delta T$ reaches zero. The left endpoint of the fitting interval, $t_0$, has
been progressively increased. The typical result of the two fitting procedure
for the GB calculation is reported in Fig. \ref{fig:Taufit}. In particular, the
two decay times for the bi-exponential fit and the single decay time for the
mono-exponential fit are plotted as a function of $t_0$. Up to approximately 20
ps, a bi-exponential behavior has been recovered with a leading decay time
$\tau_1$ of 126.5 ps and a second one, $\tau_2$, 15 to 20 times
smaller. After this transient, the second decay time goes to zero and the
expected mono-exponential behavior is reached. Consistently, the
mono-exponential fit shows a decay time coinciding asymptotically
with $\tau_1$.
Such a result is in agreement with the analysis of the heat equation solution
reported in the previous section and provides a robust fitting procedure to
extract the leading decay time from the AEMD simulation.

In order to evaluate the interface resistance from this leading decay time
via Eq. (\ref{eq:r1r2}), the bulk conductivity $k$ is needed.
As a matter of fact, at the submicrometric length scale accessible by molecular
dynamics, the silicon thermal conductivity depends on the
system length. Since the system size is comparable to the phonon mean free
paths, the dynamics of a part of the heat carriers is indeed non-diffusive, which affects
the effective thermal conductivity. 
Therefore, the value of $k$ in Eq. (\ref{eq:r1r2}) has to be carefully defined.
This could be a purely technical difficulty in the present calculation of the
interface resistance since for a material with shorter phonon mean free path
such a problem does not occur. However, this issue goes beyond the mere
technicality of the choice of the materials, and rather it may represent
a crucial point for all studies of thermal properties at the nanoscale based on
the classical heat equation. 
For this reason, in the next section we carefully discuss the size dependence 
of the bulk conductivity obtained via AEMD.

\section{Non-local formulation of the heat transfer}\label{sec:NonLocalK}

\begin{figure}[ht]
\begin{center}
\includegraphics[width=80mm]{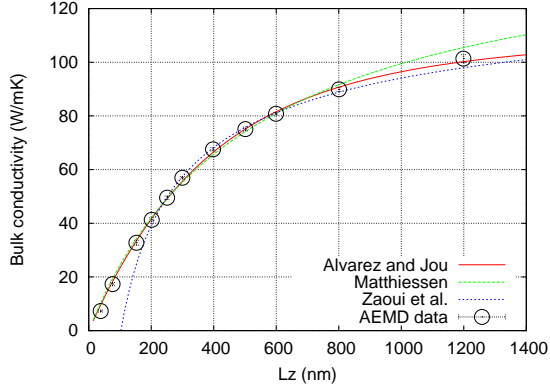}
\end{center}
\caption{\label{fig:KvsLSi} c-Si bulk conductivity at 500 K, obtained via AEMD 
simulations \cite{refJAPLampinPalla}, plotted as a function of the simulation cell length $L$. 
The data are fitted by three different analytical models.}
\end{figure}

In Fig. \ref{fig:KvsLSi}, we show the AEMD results for the silicon thermal
conductivity of our previous work \cite{refJAPLampinPalla}. We have shown
that for system sizes in the range [30 nm, 1200 nm] $k$ strongly depends on
the length $L$ of the simulation cell.
Such an effect is also recovered in MD calculations performed with different
techniques. A distinct length-scale dependence is obtained in fact both with
non-equilibrium MD approaches (NEMD) and via equilibrium 
simulations (Green-Kubo). Moreover, in these cases the effective definition of the
length-scale parameter is often uncertain \cite{Howell1}. In the NEMD
calculations a thermal gradient is induced among a heat source and a heat sink.
In between, a region with a linear gradient is identified so to calculate the
conductivity as the ratio of the heat flux to the temperature derivative. The
result depends on the distance of the source from the sink, but it is not
clear if the effective length is the depth of the linear region or rather
the source/sink distance.

A similar question arise in AEMD. Does the value of $k$ depend on
the simulation cell size $L$ (so it can be considered an artifact of the PBC)
or it depends on the wavelength $\lambda$ of the TG?
In order to answer the question, in table \ref{table:LzLamk} we report the
values of $k$ obtained varying both $L$ and $\lambda$ according to $L = n
\lambda$, $n=1,2,3$. 
\begin{center}
\begin{table}
\caption{\label{table:LzLamk}  c-Si bulk conductivity at 500 K calculated via AEMD simulations for different values of the 
length $L$ of the simulation cell (i.e. of the PBC period) and of the wavelength $\lambda$ of TG.
}
\begin{tabular}{c c c}  
 \hline
 $\lambda$ (nm) & $L$ (nm) & $k$ (W/m/K) \\ [0.5ex] 
 \hline\hline
 87.3 & 87.3 & 22  \\ 
 87.3 & 174.6 & 20  \\
 87.3 & 261.9 & 21 \\
 \hline
 174.4 & 174.4 & 36  \\
 174.4 & 348.8 & 38  \\ 
 \hline
 250 & 250 & 54   \\
 250 & 500&  52 \\ 
\hline
 350 & 350 &  64 \\
 350 & 700 &  64 \\ 
\hline
\hline
\end{tabular}
\end{table}
\end{center}
These results clearly prove that the value of $k$ obtained via AEMD depends on the wavelength of the TG $\lambda$.

A wavelength dependence of transport parameters at the nanoscale represents a
typical deviation from the classical transport theory and it is usually addressed by
considering a non-local transport equation. 
To this aim, a wavenumber dependent conductivity is introduced in the 
Fourier transform of Eq. (\ref{eq:Fourier})
\begin{equation}
\tilde J(\alpha,t) = -i  {\alpha} \tilde k(\alpha) \tilde T
(\alpha,t) \label{eq:Fouriergen}
\end{equation}
where $\alpha=\frac{2 \pi}{\lambda}$ is the wavenumber and tilde symbol
stands for the Fourier transform
\begin{equation}
 \tilde f(\alpha) = \int_{-\infty}^{\infty} f(z)e^{-i \alpha z}  dz
\end{equation}

In the macroscopic limit $\lambda \to \infty$, $\tilde k(\alpha)$ tends to the 
 macroscopic conductivity $k^\infty$.
In the real space, Eq. (\ref{eq:Fouriergen}) reads
\begin{equation}
J(z,t) =  -\int_{-\infty}^{\infty}
k(\xi) \frac{\partial T(z-\xi,t)}{\partial z} d\xi
\label{eq:Fouriernonloc}
\end{equation}
where $k(z)$ is an even function representing the non-locality of the thermal transport at 
the microscopic scale and tents to zero at some distance $\Delta z$ from origin. 
Therefore, the property $\Delta z$ turns to be the actual range of the non-locality. 
In other words, Eq. (\ref{eq:Fouriernonloc}) states that the heat flux in a given point $z$ 
depends on the temperature gradient in the neighborhood $(z-\Delta z, z+\Delta z)$ while, in 
the local formulation, it depends only on the gradient in z.
In the macroscopic limit, $\Delta z$ is negligible, hence $k(z)$ tents to $k^\infty \delta(z)$, being
$\delta(z)$ the Dirac delta function, and the local Fourier's law is recovered.
Eq. (\ref{eq:Fouriergen}) leads to the following non-local heat equation
\begin{equation}
c  \frac{\partial T(z,t)}{\partial t} =  \int_{-\infty}^{\infty}
k(\xi) \frac{\partial^2 T(z-\xi,t)}{\partial z^2} d\xi
\label{eq:heateqnonloc}
\end{equation}

As a consequence of last considerations, the curve in Fig. \ref{fig:KvsLSi}
is naturally interpreted in this contest as the Fourier
transform $\tilde k(\alpha)$ introduced in Eq. (\ref{eq:Fouriergen}). 

As reported in the previous section and largely discussed in literature \cite{sellan,ZaouiPRB},
 the scale dependence of $k$ is considered a manifestation of the non-diffusive
behavior of heat carriers with a long phonon mean free path 
and the $k(L)$ curve is usually fitted by the following formula based on the
Matthiessen's rule approximation
\begin{equation}
 \frac{1}{k(L)} = \frac{1}{k^\infty} \left(1+\frac{L_0}{L}\right)
\label{matthiessen}
\end{equation}
where $k^\infty$ is the asymptotic (macroscopic) conductivity and $L_0$ a
parameter related to the average phonon mean free path.
Eq. (\ref{matthiessen}) is widely adopted to extract the macroscopic
conductivity from MD calculations but usually the results strongly depend on
the fitting range, at least for materials as crystalline silicon which exhibit
long phonon mean free paths.
In a precedent paper \cite{ZaouiPRB}, we proved that the application of this model in AEMD is not 
fully satisfactory and we proposed a different approximation that better matches
the AEMD results at large system sizes, namely:
\begin{equation}
 {k(L)} = {k^\infty} \left(1+\sqrt{\frac{\Lambda_0}{L}}\right)
\label{zaoui}
\end{equation}
While this formula has the advantage of providing a very accurate modeling of
the values of $k$ at large $L$, it is however still unsatisfactory
since it identifies the length dependence with a merely numerical artifact,
namely the size of the simulation supercell.

On the other hand, with the present interpretation of the scale effect in terms of wavelength dependence,
a new analytical model can be invoked. Alvarez and
Jou \cite{Alvarez} indeed, starting from Eq. (\ref{eq:Fouriernonloc}) and
applying Boltzmann equation, state the following formula for $\tilde k(\lambda)$
\begin{equation}
\tilde k (\lambda) = \frac{{k^\infty \lambda^2}}{2 \pi^2 \lambda_0^2} \left[
\sqrt{1+ 4\left(\frac{\pi \lambda_0}{\lambda}\right)^2} -1 \right]
\label{alvarez}
\end{equation}

In Fig. \ref{fig:KvsLSi},  the AEMD results are fitted with the three models
described above. The Alvarez formula perfectly fits the simulation data on the
entire range. The numerical results of the fits of the AEMD data are reported
in Table \ref{table:kinf}. 
\begin{center}
\begin{table}
\caption{\label{table:kinf} Results of the fits of the c-Si bulk 
conductivity, obtained via AEMD simulations, for different analytical models.
}
\begin{tabular}{c c c}  
 \hline
 model & $k^{\infty}$ (W/mK) & $L_0$, $\Lambda_0$, $\lambda_0$ (nm) \\
[0.5ex] 
 \hline\hline
 Mathiessen, Eq. (\ref{matthiessen}) & 154 $\pm$ 6   & 520 $\pm$ 50  \\
 \hline
 Zaoui et al., Eq. (\ref{zaoui})     & 140 $\pm$ 2 & 103 $\pm$ 2  \\
 \hline
 Alvarez, Eq. (\ref{alvarez})      & 110 $\pm$ 3 & 135 $\pm$ 7  \\
 \hline
\end{tabular}
\end{table}
\end{center}
The macroscopic conductivity $k^{\infty}$ largely depends on the considered
model. We remark that a value of c-Si macroscopic conductivity at
500 K of 154 W/mK coincides with the results usually obtained in
MD \cite{HowelJCP} by means of other techniques but still extrapolated via the
Mathiessen formula. On this basis the interatomic potential here adopted, namely
the Tersoff potential (see \ref{app:simdetails}), is considered to
significantly overestimate the thermal conductivity (experimentally of 80
W/mK at 500 K).
We underline that by means of Alvarez formula the macroscopic conductivity
 predicted by AEMD (110 W/mK) is closer to the experimental value than
the other estimations. 
This means that probably the Tersoff potential is not so inadequate to a
quantitative study of the thermal transport in silicon.
Moreover, important variations of the extrapolated value can be obtained by
varying the range of the fit, namely of
22\% for the Matthiessen formula, of 19\% for the Zaoui et al. and 8\% for
Alvarez formula.

\begin{figure}[h]
\begin{center}
\includegraphics[width=80mm]{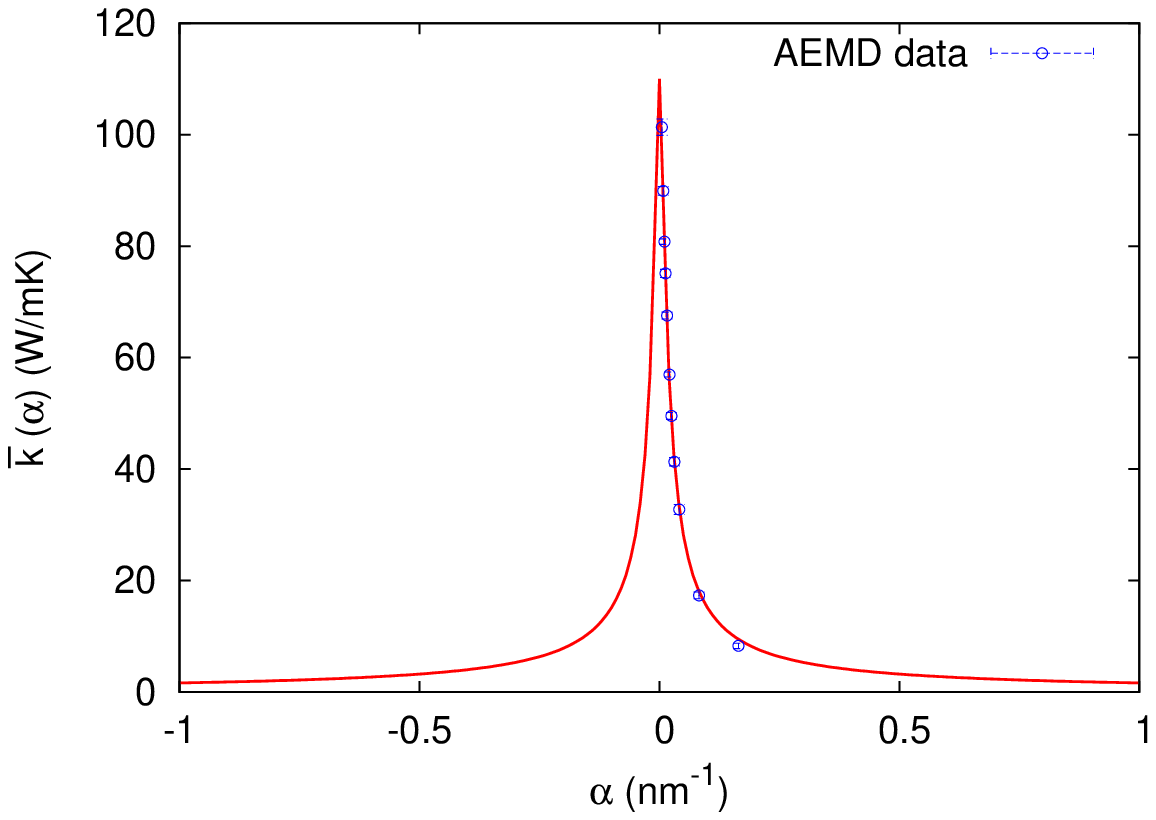}
\includegraphics[width=80mm]{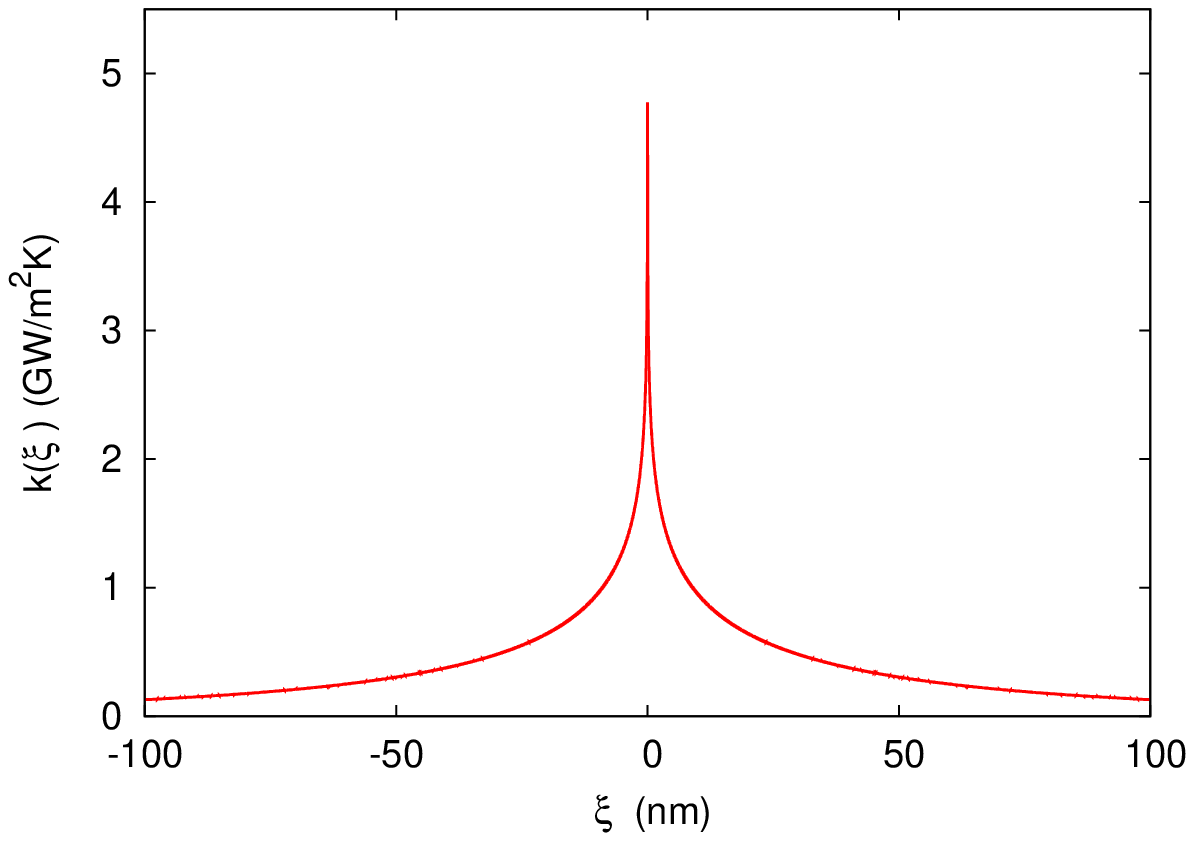}
\end{center}
\caption{\label{fig:NonLinKzalpha} Non-local thermal conductivity $\tilde k (\alpha)$ 
calculated in AEMD simulations (top panel), the continuous line has been obtained by 
fitting the data on the Alvarez formula in Eq. (\ref{alvarez}). In the bottom panel, we 
report the corresponding Fourier transform
. }
\end{figure}

The present results show that the Alvarez formula, developed in the contest of
non-local thermal transport, seems to be the best suited to model the AEMD
data.
This confirms one of the main results of this work: the thermal
transport in dielectric materials at the nanoscale can be modeled by the
classical heat equation providing that a non-local formulation is considered.
The non-local thermal conductivity, both in $\alpha$-space and in $z$-space are
shown in top panel and in bottom panel of Fig. \ref{fig:NonLinKzalpha}
respectively. The $\tilde k (\alpha)$ curve has been obtained by means of 
Eq. (\ref{alvarez}) with the parameters in Table \ref{table:kinf}. The
corresponding curve in the $z$-space, obtained by numerical integration, 
shows that the range of the non-locality $\Delta z$ is approximately 50 nm.

Actually, the non-local formulation of the heat transfer induces a different analytical modeling of
the AEMD simulations. Namely, in Section \ref{secHeatequation}, Eqs. (\ref{eq:heateq}) and (\ref{eq:Fourier}) 
 have to be replaced by Eqs. (\ref{eq:heateqnonloc}) and (\ref{eq:Fouriernonloc}),
respectively. In \ref{app:nonlocal}, we show that in spite 
of the modification of the analytical model, the main results of Section \ref{secHeatequation}
are still valid if the constant bulk conductivity $k$ is replaced by the its non-local
counterpart $\tilde k(\alpha)$.

\section{AEMD calculation of the grain boundary
thermal resistance}\label{sec:resfromtau}

In Section \ref{AEMDresistance}, we shown how to extract the leading decay
time $\tau'_1$ from AEMD simulations of a bilayer system of periodicity $L$.
In order to calculate the corresponding interface thermal resistance $r$ 
via Eq. (\ref{eq:r1r2}), the value of the bulk conductivity is needed.
In the above Section, we have seen that the bulk conductivity is actually a
function of the wavenumber $\alpha$ of the TG. As a consequence, in the
precedent analytical results, $k$ must be replaced by $\tilde k(\alpha)$.
In particular, Eq. (\ref{eq:r1r2}) reads
\begin{eqnarray}
r = 2\frac{ \sin(2l\alpha) + \sqrt{ 2(1-\cos(2l\alpha))
}}{(1-\cos(2l\alpha)) \tilde k(\alpha) \alpha }\\
\label{eq:r1r2nonloc}
\end{eqnarray}
where $\tilde k (\alpha)$ is given by the Alvarez formula 
\begin{equation}
\tilde k (\alpha) = \frac{2 k^\infty}{ (\alpha \lambda_0)^2} \left[\sqrt{1+
\left(\alpha \lambda_0\right)^2} -1 \right]\label{alvarezalpha}
\end{equation}
with parameters fitted on the AEMD data for the bulk system.
The value of $\alpha$ is obtained from the $\tau'_1$  
according to
\begin{equation}
 \tau'_1 = \frac{c}{\tilde k (\alpha)\alpha^2 } \label{eq:taualphanonloc}
\end{equation}
i.e. to 
\begin{equation}
\alpha = \frac{c}{k^\infty \tau'_1} \left( 1 + \frac{\lambda_0^2 c}{4 k
^\infty \tau'_1} \right)
\label{eq:taualphanonlocAlv}
\end{equation}
if $\tilde k(\alpha)$ in Eq. (\ref{alvarezalpha}) is assumed.

In other words, in the non-local formulation of the heat transfer, the effective 
bulk conductivity of a bilayer system of periodicity $L$ is given by the
conductivity at the wavelength $\lambda_1=\frac{2 \pi}{\alpha_1}$ (greater
than $L$) solution of Eq. (\ref{eq:det}) and related to the decay time $\tau'_1$
by Eq. (\ref{eq:taualphanonloc}).

An empirical proof of the coherence of this statement can be obtained by
calculating the values of $k$ by Eq. (\ref{eq:taualphanonloc}), i.e.
\begin{equation}
k= \frac{c \lambda^2}{\tau'_1 4\pi^2 } \label{eq:kfromlambdafit}
\end{equation}
with both $\tau'_1$ and $\lambda$ extracted from the MD simulations of the
bilayer system. Indeed, the wavenumber, and therefore $\lambda$, can
be directly extracted from the simulated TG (see Fig. \ref{fig:tempprof}).
In Table \ref{tab:lambdafits}, we report the results of this analysis. For
different values of the system size $L$, we show the corresponding decay time
$\tau'_1$ and wavelength $\lambda$ deduced from the simulations. 
The effective value of $k$ is then calculated via Eq. (\ref{eq:kfromlambdafit}).
Finally, in the Table we show the values of $k(\lambda)$ obtained in the bulk
simulations.
\begin{center}
\begin{table}
\caption{\label{tab:lambdafits} Values of the decay time $\tau'_1$ and of the wavelength $\lambda$
of the temperature profile extracted from numerical simulations of the bilayer system
(see bottom panel of Fig. \ref{fig:tempprof}) for different values of the TG period $L$.
}
\begin{tabular}{c c c c c}  
 \hline
$L$  & $\tau'_1$ & $\lambda$  &  $k$ via Eq.(\ref{eq:kfromlambdafit}) &
$k(\lambda)$ via Eq.(\ref{alvarezalpha}) \\ [0.5ex] 
(nm) &  (ps)&  (nm) &  (W/m/K)  &  (W/m/K)  \\  
 \hline\hline
32.7  &    25     &   90        &    17    & 26 \\
87.3  &    48     &   180       &    35    & 41 \\
150   &    77     &   285       &    55    & 55 \\
250   &    126    &   390       &    64    & 68 \\
400   &    218    &   575       &    79    & 81 \\
600   &    798    &   790       &    90    & 88 \\
\hline
 \hline
\end{tabular}
\end{table}
\end{center}
These data show that the effective conductivity $\tilde k(\alpha)$ calculated
via Eq. (\ref{eq:kfromlambdafit}) with $\tau'_1$ and $\lambda$ extracted from
the AEMD simulation of the bilayer system coincides with the conductivity
of the bulk at this wavelength $\lambda$ (minor discrepancies are due to the
rough estimation of $\lambda$ supplied by the fitting of the noisy and time
depend temperature profile obtained in the simulations).
This analysis substantiates the application of the non-local heat equation in the
numerical evaluation of the ITR at the nanoscale.

\begin{figure}[h]
\begin{center}
\includegraphics[width=80mm]{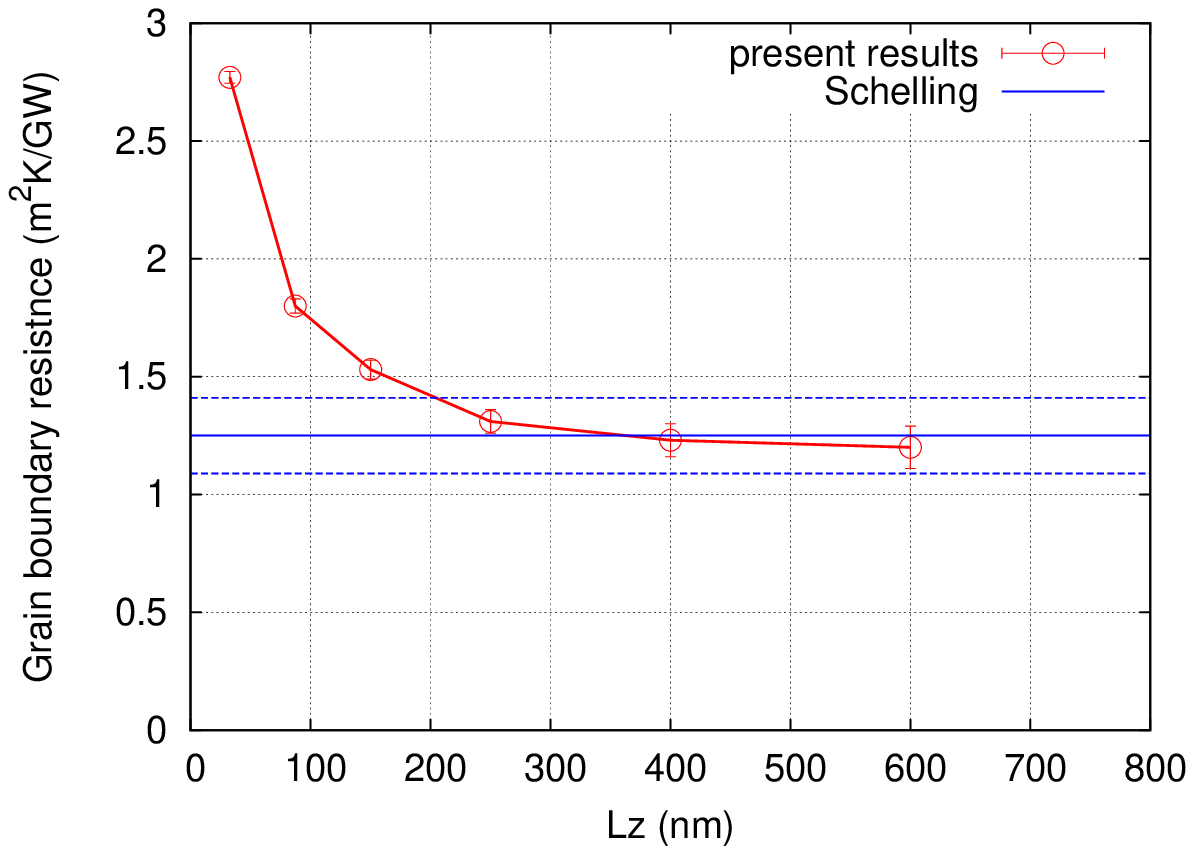}
\end{center}
\caption{\label{fig:resistances} GB thermal resistance calculated via AEMD method.
}
\end{figure}

In Fig. \ref{fig:resistances}, we plot the
ITR of the c-si GB calculated via Eqs. (\ref{eq:kfromlambdafit}),
(\ref{alvarezalpha}), and (\ref{eq:taualphanonlocAlv}) with the values of $\tau'_1$
deduced from AEMD simulations of bilayer systems of different periodicity $L$.
A size dependence of the ITR is observed up to approximately 200 nm.
On the other hand, the asymptotic value (1.2 m$^2$K/GW) is in good agreement
with the value obtained by Schelling {\it et al.} \cite{Schelling-rGBsilicon}
and by other authors \cite{twistsResistances} by means of the NEMD technique.

The interface thermal resistance is an inherently local
property, therefore a scale effect is in principle unexpected.
Other works \cite{finitesizereseffect1,finitesizereseffect2} have found
such a scale dependence in ITR numerical calculations. This artifact has been 
associated to the effects of the ballistic phonon transport.
As discussed in Section \ref{sec:NonLocalK}, in the present formulation non-diffusive effects 
are taken into account by the non-local modeling of the heat transfer of the bulk. 
Nevertheless, the non-local conductivity $k(\xi)$, here
adopted for the interpretation of the heat transfer in presence of ITRs,
doesn't account for the presence of these inhomogeneities.
The $k(\xi)$ curve, shown in the bottom panel of Fig.
\ref{fig:NonLinKzalpha}, represents indeed an homogeneous
non-locality relating the heat flux in a given point to the derivative of
temperature at a distance $\xi$ (see Eq. (\ref{eq:Fouriergen}). 
When an interface is present at this distance, the smooth $k(\xi)$ function is no
longer appropriate. 
A more advanced model could in principle be conceived with a more general
non homogeneous kernel function $k(z,z')$, in spite of the present
homogeneous kernel $k(z'-z)$.
Nevertheless, in Fig. \ref{fig:NonLinKzalpha}, we can observe that the range of
the non-locality, $\Delta z$, is around 50 nm. Therefore, the present homogeneous
approximation is admissible when the distance between the interfaces
sufficiently exceeds this limit. As a matter of fact, beyond 200 nm the
obtained value of the ITR is constant and it coincides with the result available in literature.
This means that, above this threshold, the homogeneous kernel correctly describes
the heat transport in this nanostructure. 
We remark that, up to at least 1200 nm, the ballistic effects
are still prominent (see Fig. \ref{fig:KvsLSi}). Hence, we state that in this 
range the present non-local formulation is able to model the ballistic effects
in bilayer silicon systems.

In this section, we shown that AEMD technique can be efficiently applied to the numerical evaluation of ITRs, 
also in presence of non diffusive effects. 
In the contest of the atomistic simulation, several methods exist to
study the thermal behavior of the interfaces. Nevertheless, they often
assume the lamped approximation, namely a constant temperature profile into
the bulk regions. 
The AEMD method doesn't require this approximation and, exploiting transient phenomena,
it is a time-saving approach able to address the interface thermal conduction at the nanoscale.

\section{Experimental application}\label{sec:phase}

The experimental study of the thermal transport in nanostructures with possible 
observations of non-diffusive phenomena, is a very challenging task.  
Several techniques are usually applied as the Raman thermometry \cite{expRaman}, 
3$\omega$ method \cite{exp3omega1,exp3omega2}, time-domain
thermo-reflectance \cite{expTDTR} or scanning thermal microscopy. 
Nevertheless, important limitations are however present. These methods often
work in the optical regime with a consequently reduced spatial resolution.
Device-based approach or other methods which adopt metallic heaters in thermal
measurements, couple the sample to extraneous structures. This induces
additional difficulties in the extrapolation of the thermal properties of the
sample from the response of the overall system, especially in the case of non-diffusive transport.

On the other hand, the LITG technique offers a non-invasive approach to the
thermal studies allowing for a high accuracy and reproducibility of the results.
Moreover, the thermal response is obtained as a function of wavelength
of the TG, i.e. the period of the grating, therefore non-local
effects can be directly detected. Indeed, by means of this experimental
technique, a non-diffusive behavior revealed by a grating wavelength dependence
has been already verified in silicon membranes
\cite{PRL-Nelson, Vega-LI-TTG-advances}.
In this case the scale effect is due to the nanometric thickness of the
structure and it is obtained for grating periods in the micrometer scale. 

In the above Sections, we showed how to apply the present model in AEMD simulations 
in order to predict the thermal behavior of interfaces.
Since the AEMD is the straightforward numerical counterpart of the LITG technique,
a similar setup could be implemented in this experimental approach, opening 
the way to a direct measurement of interfaces thermal resistances.


One of the first difficulties in the experimental implementation of the present model is 
the generation of TG with a sufficiently short wavelengths.  
In order to obtain a significant thermal effect of the interfaces, indeed, the spatial
period of the structure, and consequently of the thermal grating, have to be
comparable to the Kapitza length $rk$ of the interface. For silicon grain boundaries, 
the present work shows that the Kapitza length is about 100 nm.
Actually, the standard LITG techniques are performed in the optical regime,
therefore the wavelengths are limited to the microscale. Nevertheless, recent
advances in laser physics based on four-wave mixing processes allow for shorter 
wavelengths thermal gratings, i.e. in the extreme ultraviolet and soft-X-ray ranges \cite{NatureNanoLITG}.

\begin{figure}[ht]
\begin{center}
\includegraphics[width=80mm]{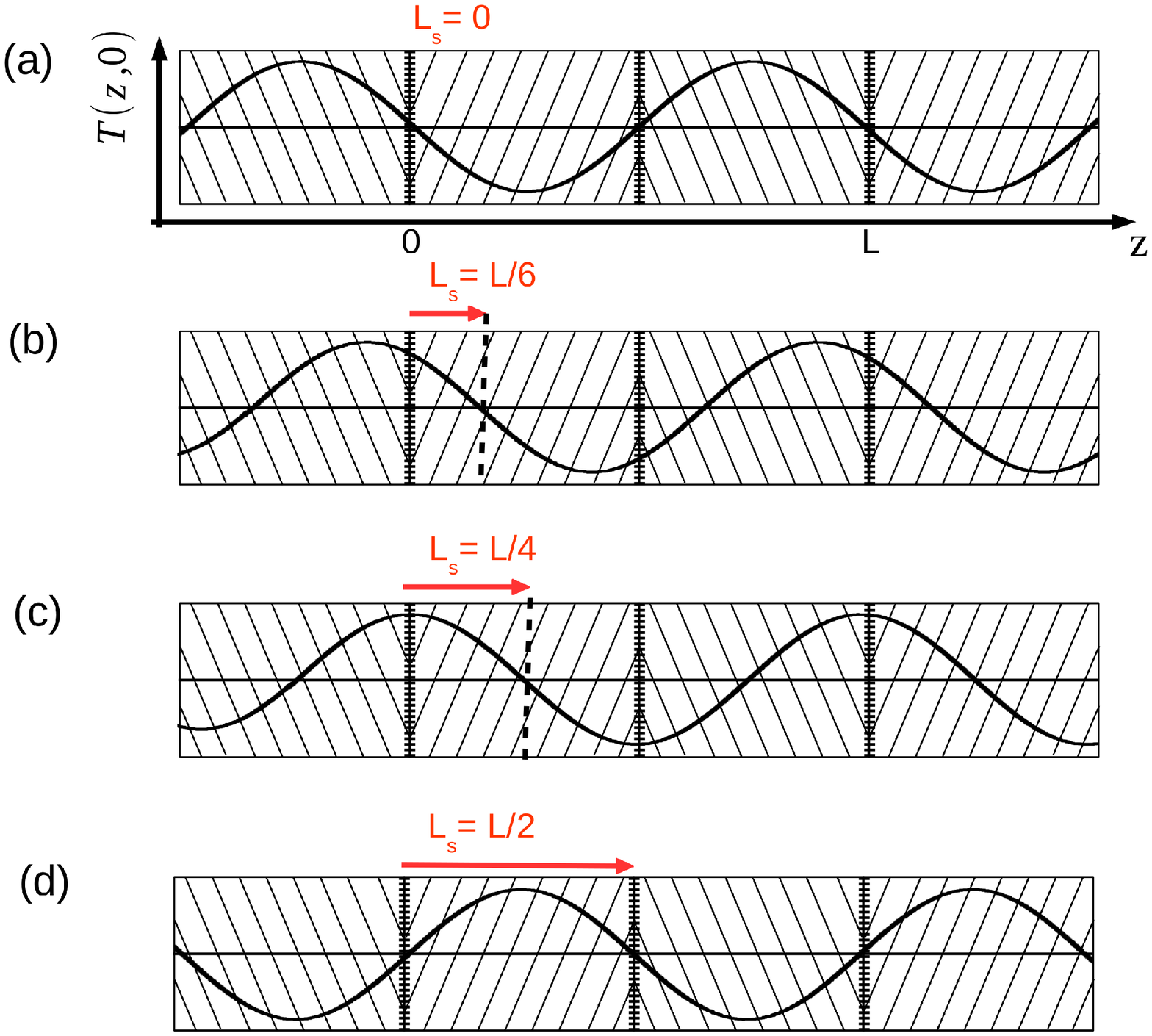}
\end{center}
\caption{\label{fig:phases} Different phases of the TG with respect to the periodic structure of the system. 
The parameter $L_s$ represents the relative shift: $L_s=0$ (panel a) corresponds to the in-phase configuration,
$L_s=\frac{L}{4}$ (panel c) to the the quadrature configuration. For $L_s=\frac{L}{2}$ the configuration is equivalent to the in-phase one.}
\end{figure}

Furthermore, the interface resistance calculation strongly depends on the knowledge of the 
conductivity of the surrounding bulk. Therefore, an accurate experimental determination of
the effective $k$ in the bulk material is required.
To address this last issue, we study here the effect of a possible
phase shift between the periodic multilayer structure and the laser induced
TG (see Fig. \ref{fig:phases}). We show that a variable phase shift can be
exploited to define a unified measurement protocol for both conductivities and
interface thermal resistance. 

\begin{figure}[ht]
\begin{center}
\includegraphics[width=80mm]{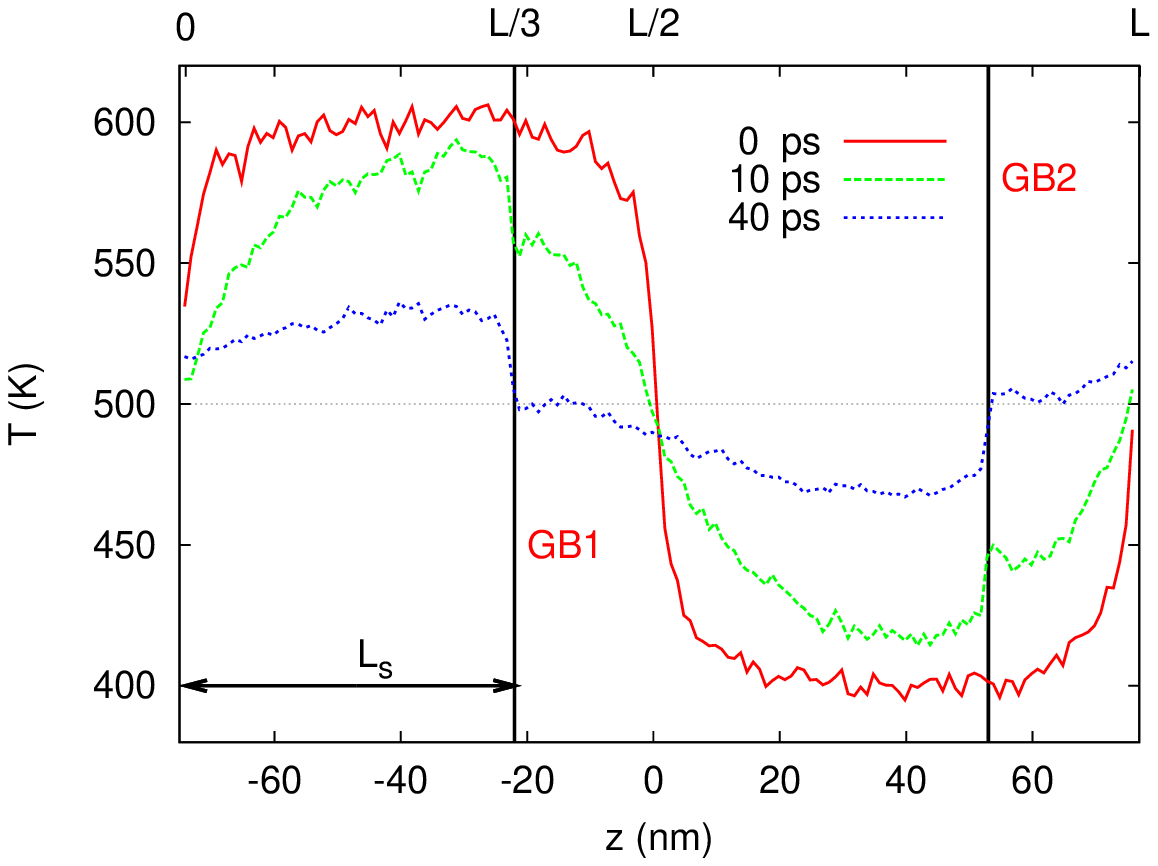}
\includegraphics[width=80mm]{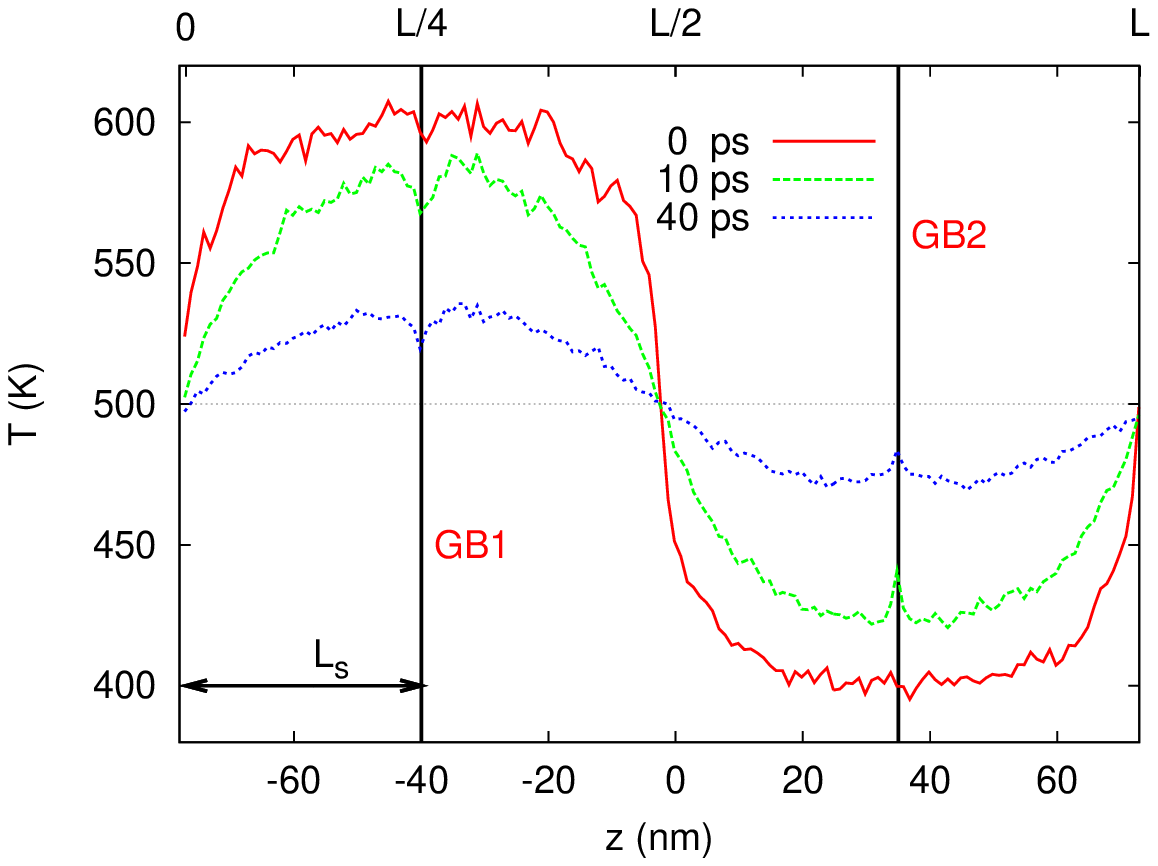}
\end{center}
\caption{\label{fig:phaseProfiles} Relaxation of the temperature profile in
the bilayer system in case of a phase shift between the TG and the periodic
structure.
}
\end{figure}

We performed AEMD simulations on the bilayer system by considering a
shift $L_s$ of the temperature profile with respect to grain boundaries positions, as
shown in Fig. \ref{fig:phases}. 
When $L_s=0$, the GBs lies on the zeros of the sinusoidal temperature profile and the TG is
an odd function. This is the \textit{in-phase} configuration adopted so far.
If $L_s = \frac{L}{4}$, the GBs lies on the maximum/minimum of the profile and
the TG is an even function in \textit{quadrature} with respect to the periodic structure. 
Finally, if $L_s = \frac{L}{2}$, the configuration is equivalent to the in-phase
case being just the cold and hot regions exchanged.
The in-phase configuration has been discussed in the previous Sections, the
leading decay time is $\tau'_1$ and it is related to the ITR by
Eq. (\ref{eq:r1r2}).
Conversely, in the quadrature configuration the heat flux at the GBs is zero and
hence there is no discontinuity in the temperature profile (see Fig. 
\ref{fig:phaseProfiles}). As a consequence, in this particular case, the
ITR doesn't have any effect on the relaxation and the bulk leading decay time
$\tau^0_1$ have to be recovered. This consideration has been mentioned in Section
\ref{secHeatequationMultiL} in order to discuss the two sequences of solutions,
$\{\alpha^0_n\}$ and $\{\alpha'_n\}$, of Eq. (\ref{eq:det}).

\begin{figure}[ht]
\begin{center}
\includegraphics[width=80mm]{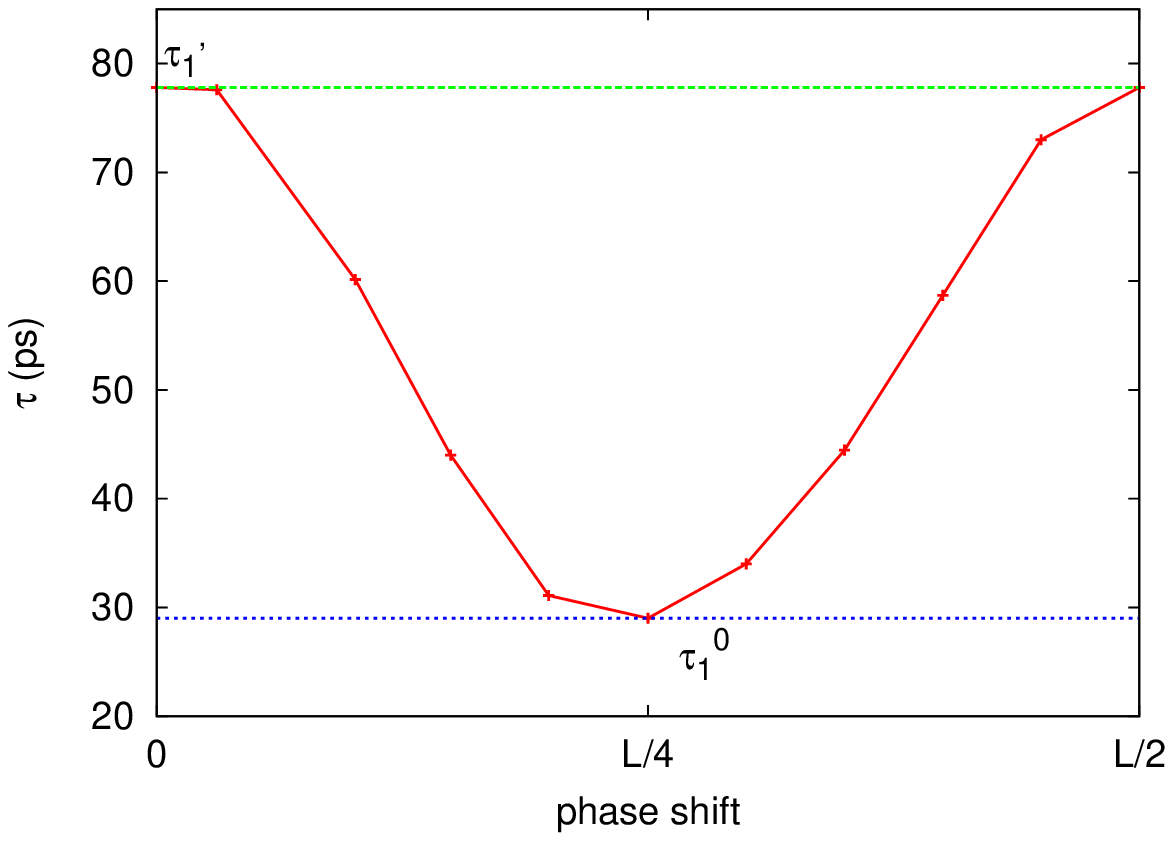}
\includegraphics[width=80mm]{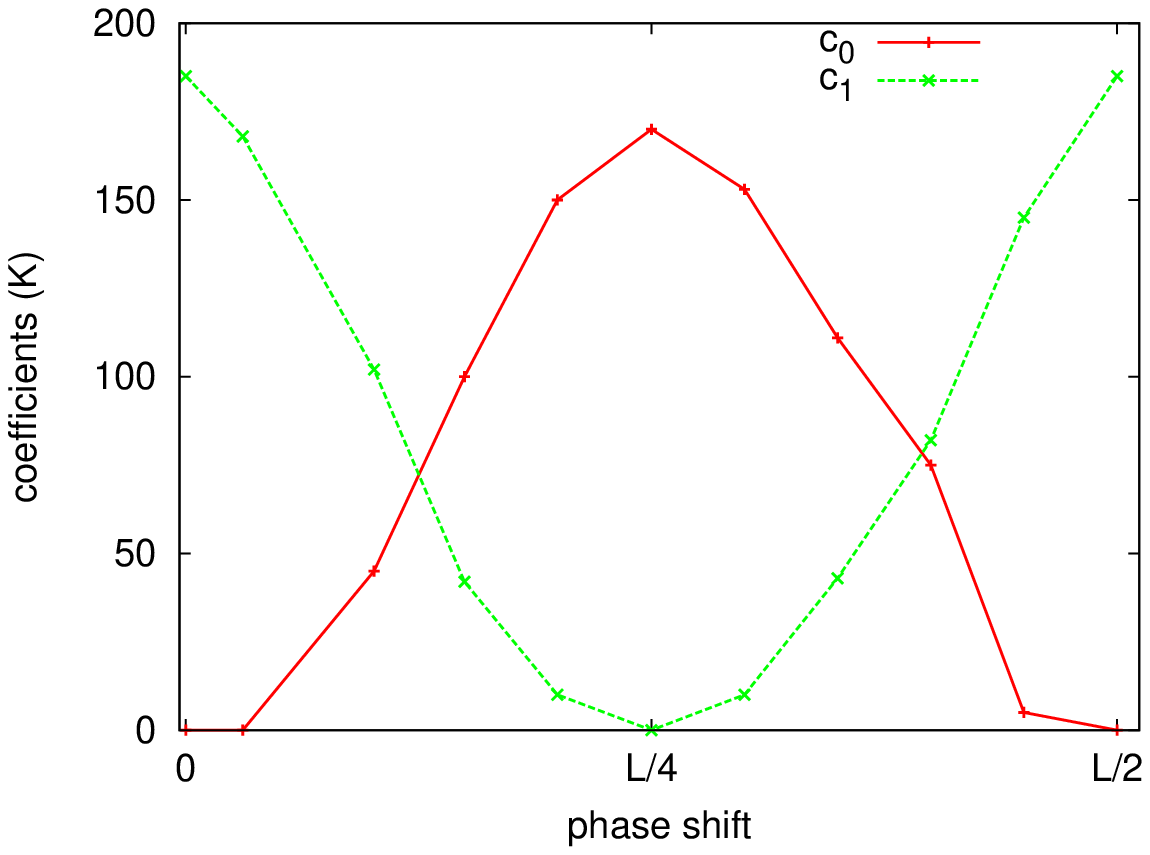}
\end{center}
\caption{\label{fig:phaseResults} Leading decay time (top panel) and coefficients (bottom panel) in Eq. (\ref{eq:fitphase})
as function of the phase shift. The data have been obtain by fits of the simulations of a bilayer silicon system of periodicity $L=250$ nm. 
}
\end{figure}

In Fig. \ref{fig:phaseResults}, we report the analysis of the relaxation
calculated in AEMD simulations as function of the phase shift $L_s$. 
The results fully confirm the above picture. In the top panel we show the leading decay
time of the temperature difference $\Delta T(t)$ which varies in the range
$(\tau^0_1, \tau'_1)$. 
In the bottom panel of Fig. \ref{fig:phaseResults}, we show the results of the
fit of $\Delta T(t)$ curves according to
\begin{equation}
\Delta T(t)=  C_0 e^{-\frac{t}{\tau^0_1}} +  C_1 e^{-\frac{t}{\tau'_1}} \label{eq:fitphase}
\end{equation}
Consistently with the analytical model developed in Section
\ref{secHeatequationMultiL}, these results show that, when the TG and the periodic structure are in-phase,
the relaxation is mono exponential with a decay time $\tau'_1$. While, for $L_s
\in (0,\frac{L}{4})$, the relaxation is a linear combination of two components,
one with a bulk-like time decay $\tau^0_1$ and the other with the ITR-dependent
decay time $\tau'_1$. When $L_s=\frac{L}{4}$ just the bulk-like component 
is present.
With respect to experimental application, this means that a 
variable phase shift sweep the leading decay time from a {\it
minimum value} $\tau^0_1$, required to calculate $k$ through
Eq. (\ref{eq:tau0inNonhomo}), to a {\it maximum value}  $\tau'_1$, required to
obtain $r$ through Eq. (\ref{eq:r1r2}).
Therefore, by varying the phase shift the same experiment can in principle
predict the effective conductivity of the bulk regions and the interface thermal
resistance of the non-homogeneous sample.


\section{Conclusions}\label{sec:conc}

In this work, we firstly developed a generalization of the classical Sommerfeld
heat-conduction problem on a ring, by introducing non-ideal interfaces described
by Kapitza thermal resistances. The solution of this problem allows to
relate the interface thermal resistance to the decay time of a thermal
grating.
Moreover, the application of the model at the nanoscale required a non-local 
formulation of the heat transfer in order to take into account 
non-diffusive effects in materials with a long phonon mean free path.
In particular, we introduced an homogeneous non-local bulk conductivity
in the diffusive heat equation to model the scale dependence of
the thermal behavior of the bulk. This strategy allows to correctly describe the
non-diffusive effects in thermal transport at the nanoscale at least down to a lower
limit (about 200 nm for silicon).

The relaxation of a TG in a bulk material is already applied both in
numerical techniques (AEMD) and in experimental approaches (LITG) in
order to calculate the thermal conductivity.
With the present results, we extend the AEMD method to the
calculation of interface thermal resistances and
we suggest a protocol to measure this property in LITG experiments.
As a numerical case study, we have calculated the interface thermal resistances of a silicon grain boundary.
The value obtained with the present AEMD method is in agreement with the available results. 
Some aspect of the possible experimental application of the present model have been discussed 
and a protocol to obtain both bulk and interface contribution to the heat transfer in a same sample has been proposed.

Since the AEMD is the numerical counterpart of the LITG experimental technique, exactly the same setup 
can by implemented in both the approaches. Therefore, a straightforward comparison between the 
corresponding results can efficiently increase the physical understanding of the heat transport
in heterogeneous structures.
In conclusion, such a non-invasive et non-destructive LITG approach, joined to its numerical counterpart, deserves
experimental investigation for both fundamental physics and applications.


\appendix

\section{Simulation details and grain boundary models} \label{app:simdetails}
In this work, we considered the disordered (100)$\Sigma 29$
twist grain boundary in silicon. The thermal behavior of this high energy GB as been
addressed by several previous works \cite{Schelling-rGBsilicon,
Schelling-rGBcarbon}. Moreover, the average GB thermal resistance of
polycrystalline silicon evaluated via MD simulation \cite{JAP112-064305-Ju-Liang}
is very similar to that of (100)$\Sigma 29$. Therefore, this grain boundary
structure has been considered as a representative model of a generic GB in polycrystalline Si.

In all the simulations (performed with a modified version of the DL\_POLY 4
package \cite{dlpoly}), the interaction between silicon atoms has
been modeled by the Tersoff potential \cite{Tersoff}. 
The $x$ and $y$ dimensions have been fixed at 62.24 nm 
corresponding to the equilibrium lattice constants of the perfect crystals at 
the temperature $T^{eq}=500$ K. This temperature, close to the Debye temperature of silicon,
has been considered as the reference temperature for our MD calculations of the thermal
properties of this material.

In order to check the dependence of the results on the atomistic model of 
the interface, two microscopically different structures of the grain boundary have
been achieved by means of a treatment similar to that presented by Keblinski
\textit{et al} \cite{PRL77-2965-keblinski,JACS80-717-keblinski} for silicon
systems modeled by the Stillinger-Weber potential. 
  Firstly, a zero temperature equilibrium structure has been obtained by static
iterative energy minimization and then relaxed at $500$ K (\textit{ordered
GB}). 
  Then, a disordered lower energy structure (\textit{annealed GB}) has been
obtained by thermal annealing at $2000\,{\rm K}$, namely $400\,{\rm K}$ below
the melting temperature of silicon simulated via the Tersoff
potential. 
  In order to promote configurational transformations of the GB structure, during
the simulations a constant stress algorithm in the direction orthogonal to the
GB surface has been applied.

\begin{figure}
\begin{center}
\includegraphics[width=80mm]{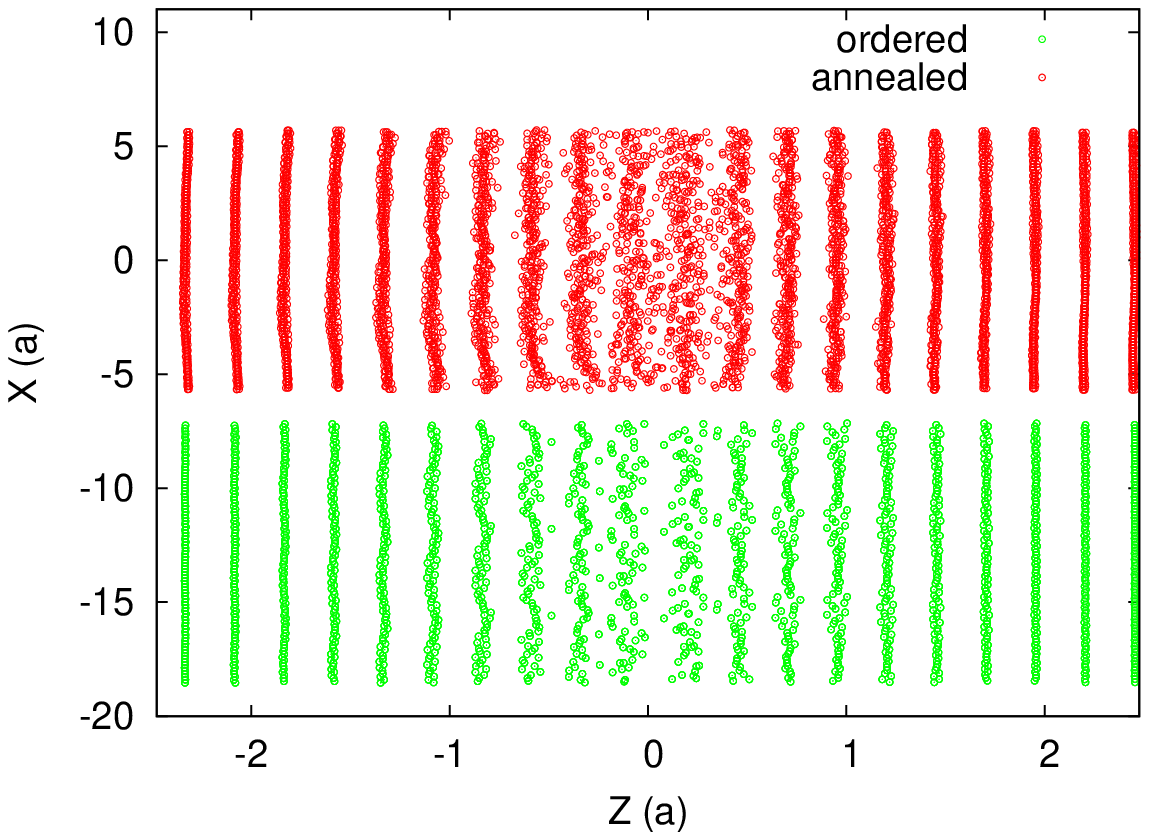}
\includegraphics[width=80mm]{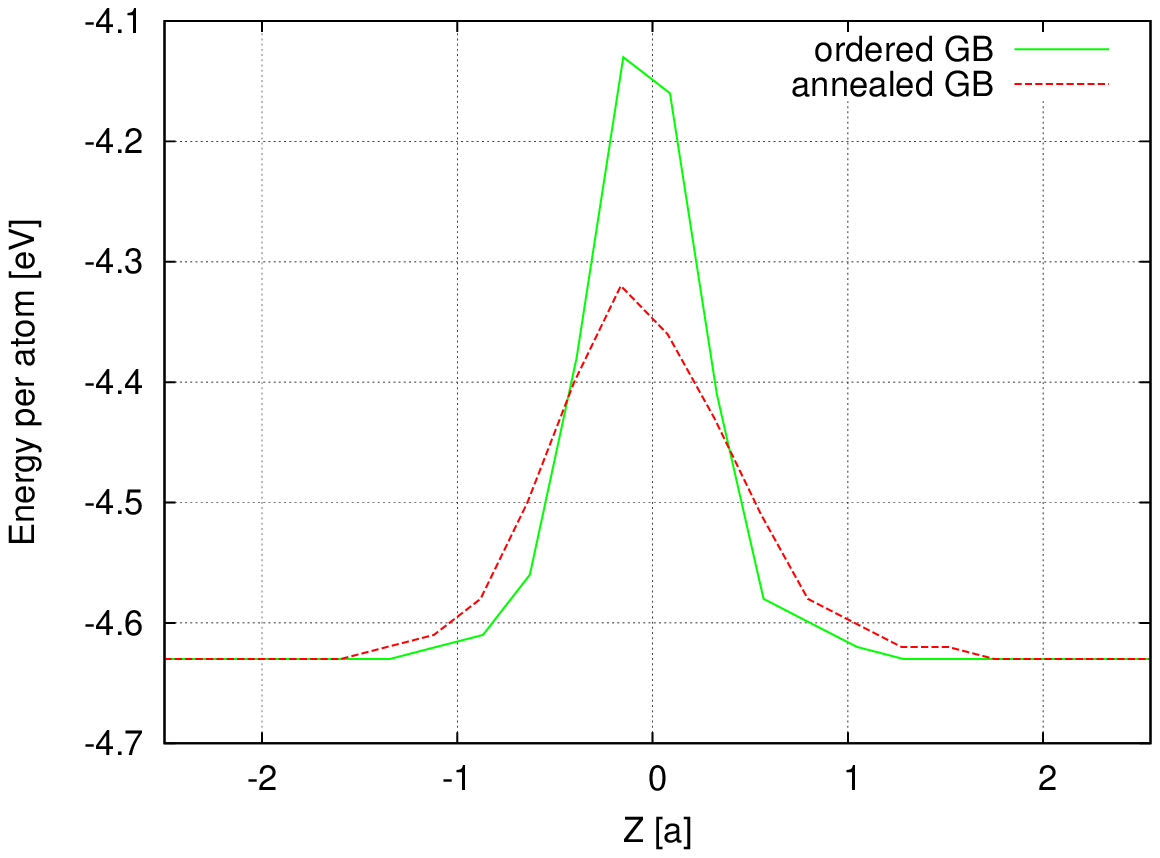}
\end{center}
\caption{\label{fig:GBenergyprof} 
  GB energy profiles and corresponding snapshots of the atomic structure at 0 K.
The z-coordinate, normal to the interfaces, is express in units of
the lattice parameter.}
\end{figure}

  For both the GB models, in Fig. \ref{fig:GBenergyprof} we plot the energy
per atom of the atomic layers parallel to the GB surface and a snapshot of the
atomic structure at $T=0$ of the two models of GB.
  The annealing process has allowed a reduction of the GB energy from
1592 erg/cm$^2$ of the ordered GB to 1409 erg/cm$^2$ of the annealed GB,
corresponding to a decrease of 11\%. These results are very similar to those
obtained by Keblinski \textit{et al} \cite{PRL77-2965-keblinski} with the
Stillinger–Weber Potential, namely 1464 \,erg/cm$^2$ and 1300-1340 \,erg/cm$^2$,
corresponding to a reduction of approximately 10\%.

In spite of the quite significant structural difference between the two models, 
the results of the AEMD calculations, reported in Table \ref{tab:GBmodelsTaus},
are not remarkably dissimilar.
\begin{center}
\begin{table}
\caption{\label{tab:GBmodelsTaus}: Leading decay times calculated via AEMD simulation of the two different models of GB.
}
\begin{tabular}{c c c }  
\hline
L (nm) & $\tau$ annealed GB (ps)& $\tau$ ordered GB (ps)  \\ [0.5ex]
\hline
\hline
32.7  &    25     &   24       \\
87.3  &    48     &   49.5     \\
150   &    77     &   76       \\
250   &    126    &   130      \\
400   &    218    &   215.5    \\
\hline
\hline
\end{tabular}
\end{table}
\end{center}

Therefore,  only the results obtained by the annealed GB model are
discussed in the paper.

\section{Non-local heat conduction problem on a ring}\label{app:nonlocal}

We consider here the non-local heat equation (\ref{eq:heateqnonloc}). The general solution can
be calculated by separation of variables. The temperature field is therefore expressed by 
\begin{equation}
T(z,t)= \chi(z)\phi(t)
\end{equation}
and Eq. (\ref{eq:heateqnonloc}) reads
\begin{equation}
 c  \frac{\partial \phi(t)}{\partial t}\frac{1}{\phi(t)} = \frac{1}{\chi(t)} \int_{-\infty}^{\infty}
k(\xi) \frac{\partial^2 \chi(z-\xi)}{\partial z^2} d\xi
\label{eq:app2.1}
\end{equation}
As a consequence, $\chi(z)$ and $\phi(t)$ verify the following equations:
\begin{eqnarray} 
\frac{\partial \phi(t)}{\partial t} &=& - \frac{\beta^2}{c}
{\phi(t)}\label{eq:app2.2} \\
 \int_{-\infty}^{\infty} k(\xi) \frac{\partial^2 \chi(z-\xi)}{\partial z^2} d\xi
&=&  -{\beta^2}{\chi(t)}
  \label{eq:app2.3}
\end{eqnarray}
where $\beta$ is a arbitrary constant.
Eq. (\ref{eq:app2.2}) states that 
\begin{equation}
\phi(t) \propto e^{-\frac{t}{\tau(\beta)}}
\end{equation}
where
\begin{equation}
\tau(\beta) = \frac{c}{\beta^2} \label{eq:taubeta}
\end{equation}
In order to solve Eq. (\ref{eq:app2.3}), we consider the Fourier transform of
$\chi(z)$ and $k(z)$ and we obtain
\begin{equation}
\tilde \chi( \alpha) \left( 1 - \frac{\alpha^2}{\beta^2} \tilde k(\alpha)\right) = 0 
\label{eq:app2.4}
\end{equation}
where $\alpha$ is the wavenumber of the Fourier transform.
$k(z)$ is an even function, hence $\tilde k(\alpha)=\tilde k(-\alpha)$ and
\begin{equation}
\tilde \chi( \alpha) = \delta\left( \alpha \pm \sqrt{  \frac{\beta^2}{\tilde
k(\alpha)} } \right) \label{eq:app2.5}
\end{equation}
Finally, we get 
\begin{equation}
\chi(z) =  C e^{i \alpha z} + C^* e^{-i \alpha z} \label{eq:gensolnonlin}
\end{equation}
where $C$ is an arbitrary complex constant and $\alpha^2 = \frac{\beta^2}{\tilde
k(\alpha)}$. Combining this last equation with Eq. (\ref{eq:taubeta}), we obtain
\begin{equation}
 \tau = {  \frac{c}{\alpha^2 \tilde k(\alpha)} } \label{eq:taualpha2}
\end{equation}
equivalent to Eq. (\ref{eq:taualpha1}).

\subsection{Periodic boundary conditions for a bulk system}

In Eq. (\ref{eq:gensolnonlin}), the allowed values of the
wavenumber $\alpha$ are defined by the boundary conditions.
We firstly take into consideration the PBC in Eqs. (\ref{pbcbulk1}) and
(\ref{pbcbulk2}) defining a periodic homogeneous system.
The non-locality of the model affects the definition of the heat
flux according to Eq. (\ref{eq:Fouriernonloc}) and leads to following set of PBC
\begin{eqnarray}
 T(0,t)&=&T(L,t)\label{eq:pbcbulk1nonloc} \\
 \int_{-\infty}^{\infty} k(\xi) \frac{\partial T(-\xi,t)}{\partial z} d\xi &=&
\nonumber \\ 
 \int_{-\infty}^{\infty} &k(\xi)& \frac{\partial T(L-\xi,t)}{\partial z} d\xi
\label{eq:pbcbulk2nonloc}
\end{eqnarray}
By replacing Eq. (\ref{eq:gensolnonlin}) in these last
equations, we find that solutions exist for all $C_1$ if $e^{i\alpha L}=1$,
hence
\begin{equation}
\alpha_n= \frac{2 \pi n }{L}
\end{equation}
Therefore, we prove that the non-local formulation just introduce the wavelength
dependent conductivity $\tilde k(\alpha)$ in the solution for the bulk case
 reported in Section \ref{secHeatequationBulk}.

\subsection{Periodic boundary conditions for a bilayer system}

We consider now the set of PBC, involving ITRs, in Eqs.
(\ref{eq:pbcGB1nonloc}), (\ref{eq:pbcGB2nonloc}), (\ref{eq:pbcGB3nonloc}), and 
(\ref{eq:pbcGB4nonloc}).
In the present non-local formulation, the heat flux reads 
\begin{eqnarray} 
J_i(z,t)= -\int_{-\infty}^{\infty} k_i(\xi) \frac{\partial T_i(z-\xi,t)}{\partial z} d\xi 
\end{eqnarray}
The PBC equations admit solution in the case $\tau^{(1)} = \tau^{(2)}$, i.e.
\begin{equation}
 {  \frac{c_1}{\alpha_1^2 \tilde k_1(\alpha_1)} }= {  \frac{c_2}{\alpha_2^2 \tilde k_2(\alpha_2)} }
\end{equation}
Moreover, the wavenumbers in Eq. (\ref{eq:gensolnonlin}) are provided, in this case, by 
 \begin{eqnarray}
&& \Psi^+ \sin \left( \alpha_1 l_1+\alpha_2 l_2\right) \\
&+&\Psi^- \sin\left( \alpha_1 l_1-\alpha_2 l_2\right) \nonumber \\
 &+&\Omega^+ \Xi^+ \cos\left( \alpha_1 l_1+\alpha_2 l_2\right)\\
 &+&\Omega^- \Xi^-\cos\left( \alpha_1 l_1-\alpha_2 l_2\right) + 4=0
 \end{eqnarray}
where
 \begin{eqnarray}
  \Psi^+ &=& 2 r \left( \alpha_1 \tilde k_1(\alpha_1)+\alpha_2 \tilde k_2(\alpha_2)\right) \\
  \Psi^- &=& 2 r \left( \alpha_1 \tilde k_1(\alpha_1)-\alpha_2 \tilde k_2(\alpha_2)\right) \\
  \Omega^+ &=&  \left( \alpha_2 \tilde k_2(\alpha_2)r + \frac{\alpha_2 \tilde k_2(\alpha_2)}{\alpha_1 \tilde k_1(\alpha_1)} +1 \right)\\ 
   \Xi^+ &=& \left( \alpha_1 \tilde k_1(\alpha_1)r - \frac{\alpha_1 \tilde k_1(\alpha_1)}{\alpha_2 \tilde k_2(\alpha_2)} -1 \right)\\
   \Omega^- &=& - \left( \alpha_2 \tilde k_2(\alpha_2)r - \frac{\alpha_2 \tilde k_2(\alpha_2)}{\alpha_1 \tilde k_1(\alpha_1)} +1 \right) \\
   \Xi^- &=&\left( \alpha_1 \tilde k_1(\alpha_1)r - \frac{\alpha_1 \tilde k_1(\alpha_1)}{\alpha_2 \tilde  k_2(\alpha_2)} +1 \right) 
   \label{eq.formulona}
 \end{eqnarray}
 If, in this last equation, we take $l_1=l_2$ and $k_1=k_2$, we recover Eq. (\ref{eq:det}) of Section \ref{secHeatequationMultiL} with the 
 non-local conductivity $\tilde k(\alpha)$ in spite of the macroscopic one.
 Therefore, also the results of Section \ref{secHeatequationMultiL} are still valid in the non-local
 formulation providing that the macroscopic bulk conductivity $k$ is replaced by its non-local counterpart $\tilde k(\alpha)$.
 Finally, Eq. (\ref{eq.formulona}) represents the generalization of Eq. (\ref{eq:det}) to the case $l_1 \neq l_2$ and $k_1 \neq k_2$.

%

\bibliographystyle{elsarticle-num}



\end{document}